\pdfoutput=1
\RequirePackage{ifpdf}
\ifpdf 
\documentclass[pdftex]{sigma}
\else
\documentclass{sigma}
\fi

\numberwithin{equation}{section}

\usepackage{cite}

\begin{document}

\allowdisplaybreaks

\renewcommand{\thefootnote}{$\star$}

\renewcommand{\PaperNumber}{096}

\FirstPageHeading

\ShortArticleName{Time-Frequency Integrals and the Stationary Phase Method}

\ArticleName{Time-Frequency Integrals and the Stationary\\
Phase Method in Problems of Waves Propagation\\from Moving Sources\footnote{This
paper is a contribution to the Special Issue ``Superintegrability, Exact Solvability, and Special Functions''.
The full collection is available at
\href{http://www.emis.de/journals/SIGMA/SESSF2012.html}{http://www.emis.de/journals/SIGMA/SESSF2012.html}}}

\Author{Gennadiy BURLAK~$^\dag$ and Vladimir RABINOVICH~$^\ddag$}

\AuthorNameForHeading{G.~Burlak and V.~Rabinovich}

\Address{$^\dag$~Centro de Investigaci\'{o}n en Ingenier\'{\i}a y Ciencias Aplicadas,
Universidad Aut\'{o}noma\\
\hphantom{$^\dag$}~del Estado de Morelos, Cuernavaca, Mor. M\'{e}xico}
\EmailD{\href{mailto:gburlak@uaem.mx}{gburlak@uaem.mx}}

\Address{$^\ddag$~National Polytechnic Institute, ESIME Zacatenco, D.F. M\'{e}xico}
\EmailD{\href{mailto:vladimir.rabinovich@gmail.com}{vladimir.rabinovich@gmail.com}}

\ArticleDates{Received July 29, 2012, in f\/inal form December 02, 2012; Published online December 10, 2012}

\Abstract{The time-frequency integrals and the two-dimensional stationary phase method
are applied to study the electromagnetic waves radiated by moving modulated
sources in dispersive media. We show that such unif\/ied approach leads to
explicit expressions for the f\/ield amplitudes and simple relations for the
f\/ield eigenfrequencies and the retardation time that become the coupled
variables. The main features of the technique are illustrated by examples of
the moving source f\/ields in the plasma and the Cherenkov radiation. It is
emphasized that the deeper insight to the wave ef\/fects in dispersive case
already requires the explicit formulation of the dispersive material model. As
the advanced application we have considered the Doppler frequency shift in a
complex single-resonant dispersive metamaterial (Lorenz) model where in some
frequency ranges the negativity of the real part of the refraction index can
be reached. We have demonstrated that in dispersive case the Doppler frequency shift
acquires a nonlinear dependence on the modulating frequency of the radiated particle.
The detailed frequency dependence of such a shift and spectral behavior of phase
and group velocities (that have the opposite directions) are studied numerically.}


\Classification{78A25; 78A35}

\section{Introduction}

The paper is devoted to applications of time-frequency integrals and the
two-dimensional stationary phase method for problems of waves propagation from
moving sources in dispersive media. We consider the electromagnetic f\/ields
generated by a moving in a dispersive media modulated source of the form
\[
\boldsymbol{F}(t,\boldsymbol{x})=\boldsymbol{a}(t)e^{-i\omega_0 t}\delta(\boldsymbol{x}-\boldsymbol{x}_0 (t)),
\qquad t\in\mathbb{R},\qquad \boldsymbol{x}=(x_{1},x_{2},x_{3})\in\mathbb{R}^{3},
\]
where $\omega_0 $ is an eigenfrequency of the source, $\boldsymbol{a}(t)$ is a
slowly varying amplitude, $\boldsymbol{x}_0 (t)=(x_{01}(t),x_{02}(t),x_{03}(t))$
is a vector-function def\/ining a motion of the source, $\delta$ is the standard
$\delta$-function.

Some assumptions with respect to sources allow us to introduce a large
dimensionless parameter $\lambda>0$ which characterizes simultaneously a
slowness of variations of amplitudes and velocities of sources, and large
distances between sources and receivers. We obtain a representation of the
f\/ields as double oscillating integrals depending on the parameter $\lambda>0$
\begin{gather}
\boldsymbol{\Phi}_{\lambda}(t,\boldsymbol{x})=\int_{{\mathbb R}\times {\mathbb R}}
\boldsymbol{F}(t,\boldsymbol{x},\omega,\tau,\lambda)e^{i\lambda S(t,\boldsymbol{x},\omega,\tau)}
{\rm d}\omega{\rm d}\tau,\label{01}
\end{gather}
where $\boldsymbol{F}$ is a complex vector-valued amplitude and~$S$ is a
real-valued for $\vert \omega\vert $ large enough phase. Generally
speaking integral~(\ref{01}) is divergent and we consider its regularization
which is called the oscillatory integral.

Applying to the integrals~(\ref{01}) the stationary phase method we obtain the
asymptotics of electromagnetic f\/ields for large $\lambda>0$. We consider non
uniformly moving sources in \textit{isotropic dispersive homogeneous media},
in a particular case, in the isotropic plasma. Note that this method can be
apply also for the analysis of waves propagation from moving sources in
anisotropic dispersive non homogeneous media and media with negative phase
velocity (metamaterials), and also for a motion with a velocity larger than a
phase velocity of media (the Cherenkov radiation).

We would like to note that the asymptotic estimates of
\textit{one-dimensional} integrals are standard tools of the electrodynamics
(see for instance \cite[Chapters~3,~4]{FM}, \cite{Felsen})
and go back to A.~Sommerfeld \cite{Som}, and L.~Brillouin \cite[Chapter~1]{Br}.
But in the case of
modulated non uniformly moving sources the representation of the f\/ields in a
form of a one-dimensional integral is impossible. In turn, a representation of
f\/ields as double time-frequency oscillating integrals with a subsequent
asymptotic analysis yields ef\/fective formulae for both the f\/ields and for the
Doppler shifts. In particular, it gives new approaches to the Cherenkov
radiation (see e.g.~\cite{Afanasiev:1998a,Afanasiev:1999,Bol,Ginz2,Ginz3,Jelly} and
\cite[Chapter~14]{LL8}). In particular, the works
\cite{Afanasiev:1998a,Afanasiev:2004a,BBM,Belonogaya:2011a,Bol,BH,Br,Carusotto:2001,
Cohen,Costas:2011a,Doilnitsina:2007a,Duan:2009a,
FM,Fedoruk,Galyamin:2010a,Ginz1,Ginz2,Ginz3,Grzegorczyk:2003a,Jackson,Jelly,Jiabi:2011a,Ilich,
Landau,LL8,Landau:1975a,Landau:1981a,Levis,Luo:2003,Muhan:2011a,Felsen,Obr1,Obr2,Obr3,
Pendry:1999a,Pendry:2006a,Press,Shubin,Som,Stix:2001b,Stix,Swen, Smith:2000a,Smolyaninov:2007a,
Shalaev:2007a,Shin:2009a,TrungDung:2003b,Tyukhtin:2008a,Tyukhtin:2005a,WH,Xi:2009a,Xiao:2010a,
Veselago:1968a,Vorobev:2012a} describe properties of the charged
particle f\/ield in dif\/ferent dispersive media including traditional resonant
medium
\cite{Afanasiev:1998a,Afanasiev:2004a,BBM,Belonogaya:2011a,Bol,BH,Br,Carusotto:2001,Cohen,
Costas:2011a,Doilnitsina:2007a,Duan:2009a,
FM,Fedoruk,Galyamin:2010a,Ginz1,Ginz2,Ginz3,Grzegorczyk:2003a,Jackson,Jelly,Jiabi:2011a,Ilich,
Landau,LL8,Landau:1975a,Landau:1981a,Levis,Luo:2003,Muhan:2011a,Felsen,Obr1,Obr2,Obr3,
Pendry:1999a,Pendry:2006a,Press,Shubin,Som,Stix:2001b,Stix,Swen,Smith:2000a,Smolyaninov:2007a,
Shalaev:2007a,Shin:2009a,TrungDung:2003b,Tyukhtin:2008a}, active
medium~\cite{Tyukhtin:2008a}, anisotropic medium~\cite{Belonogaya:2011a},
left-handed medium~\cite{Galyamin:2010a,Grzegorczyk:2003a,Duan:2009a}, and so-called
``wire-metamaterial''~\cite{Vorobev:2012a}.
Some of these works develop a method of analysis of the moving charge f\/ield
using complex function theory methods~\cite{Tyukhtin:2008a,Belonogaya:2011a}.
The papers \cite{Tyukhtin:2005a,Doilnitsina:2007a} are devoted to investigation of the f\/ields of moving
oscillators in dif\/ferent media.

The electromagnetic radiation from moving sources is a classical problem of
the electrodynamics, and for the isotropic \textit{non dispersive media} the
solution of this problem is given by the \textit{Li\`{e}nard--Wiechert
potential} (see for instance \cite[Chapter~VIII]{Landau}, \cite[Chapter~14]{Jackson}).
But the \textit{Li\`{e}nard--Wiechert potential} is not applicable for
dispersive media and our representation is an ef\/fective tool for the
investigation of electromagnetic f\/ields generated by moving sources with a
variable velocity.

Note that the above-described method for estimating of the acoustic f\/ield
generated by moving sources in \textit{stratified acoustic waveguide} has been
proposed f\/irst in \cite{Ilich}, and later on in the papers \cite{Obr1,Obr2,Obr3}.

There is another asymptotic approach to the problems of waves propagation from
moving sources in dispersive media. It is the ray method in the space of
the variables $(t,\boldsymbol{x})$ (see for instance~\cite{BBM,Levis}).
Despite the fact that the ray method is applicable to a wider
range of problems than the method suggested in the article, its implementation
is encountering very serious dif\/f\/iculties in solving the ray and transport
equations. By contrast the approach proposed here leads to simple, having a
clear physical meaning, equations for the stationary points, and explicit
formulae for electromagnetic f\/ields and Doppler shifts. In particular, in the
developed approach the distinction between the phase and group velocities
appear in a natural way.

The paper is organized as follows. In Section~\ref{section2} we give auxiliary material
concerning the oscillatory integrals and the multidimensional stationary phase
method. In Section~\ref{section3} we consider electromagnetic waves propagation from
moving modulated sources in dispersive medias. We obtain ef\/fective asymptotic
formulae for the electromagnetic f\/ields, Doppler ef\/fects, and retarded time.
Section~\ref{section4} devoted to applications obtained in Section~\ref{section4} formulae. We
consider a motion with a constant velocity in non dispersive media,
electromagnetic f\/ield generated by modulated stationary sources in dispersive
media, electromagnetic waves propagation from uniformly moving sources in the
lossless no magnetized plasma (see e.g.~\cite{Ginz3, Jackson,Stix, Swen}).
We also formulate the equations for the Cherenkov
radiation in dispersive medias in terms of representation~(\ref{01}) and the
stationary phase method.

Further it is emphasized that the deeper comprehension and insight the wave
ef\/fects in dispersive case already requires the explicit formulation of the
dispersive material model. As the advanced application of the developed
technique in Section~\ref{section5} it is considered the Doppler frequency shift in a
complex single-resonant dispersive metamaterial (Lorenz) model where in some
frequency ranges the negativity of the real part of the refraction index can
be reached~\cite{Veselago:1968a,Pendry:1999a, Smith:2000a,Smolyaninov:2007a,Shalaev:2007a}.
We have demonstrated that in dispersive case the Doppler
frequency shift acquires a nonlinear dependence on the modulating frequency
of the radiated particle. (In dispersiveless medium such a function is linear.)
The detailed frequency dependence of such shift and spectral be\-ha\-vior of phase
and group velocities (that have the opposite directions) are studied numerically.
Discussion and conclusions from our results are found in the last section.

\section{Auxiliary material: stationary phase method\\for the oscillatory
integrals}\label{section2}

\subparagraph*{$1^{0}.$}
We use the standard notations for the spaces of dif\/ferentiable
functions: $C^{\infty}(\mathbb{R}^{n})$ is the space of all inf\/initely
dif\/ferentiable functions on $\mathbb{R}^{n}$, $C_{b}^{\infty}(\mathbb{R}^{n})$ is
a subspace of $C^{\infty}(\mathbb{R}^{n})$ consisting of functions bounded
with all their partial derivatives, $C_0 ^{\infty}(\mathbb{R}^{n})$ is a
subspace of $C^{\infty}(\mathbb{R}^{n})$ consisting of functions with a
compact supports.

\subparagraph*{$2^{0}.$}
We consider integrals of the form
\begin{gather}
\int_{\mathbb{R}^{n}}\boldsymbol{f}(\boldsymbol{x})e^{iS(\boldsymbol{x})}{\rm d}\boldsymbol{x},
\label{0.0}
\end{gather}
where $\mathbb{R}^{n}\ni\boldsymbol{x}\rightarrow\boldsymbol{f}(\boldsymbol{x})\in\mathbb{C}^{m}$
is called the amplitude and the scalar function $S$ is called the phase. We
suppose that $\boldsymbol{f}$ and $S$ are inf\/initely dif\/ferentiable (the
existence only of a f\/inite number of derivatives is necessary) and satisfy the
following conditions:
for every multiindex $\alpha$ there exists $C_{\alpha}>0$ such that
\begin{gather}
\big\vert \partial^{\alpha}\boldsymbol{f}(\boldsymbol{x})\big\vert \leq C_{\alpha
} \langle \boldsymbol{x} \rangle^{k},\qquad \boldsymbol{x}\in\mathbb{R}^{n},
\qquad \langle \boldsymbol{x} \rangle =\big(1+\vert \boldsymbol{x}\vert^{2}\big)^{\frac{1}{2}}, \label{0.1}
\end{gather}
\begin{enumerate}\itemsep=0pt
\item[$(i)$] $S(x)$ is a real function for $\left\vert x\right\vert $ is large enough,
\item[$(ii)$] for every $\alpha$: $\vert \alpha\vert \geq2$ there
exists $C_{\alpha}>0$ such that $\vert \partial^{\alpha}S(\boldsymbol{x})\vert \leq C_{\alpha}$,
\item[$(iii)$] there exists $C>0$ and $\rho>0$ such that
\[
\vert \nabla S(\boldsymbol{x})\vert \geq C\vert \boldsymbol{x}\vert^{\rho}
\]
for $\left\vert\boldsymbol{x}\right\vert $ large enough.
\end{enumerate}

Note that if $k\geq-n$ integral (\ref{0.0}) does not exist as absolutely
convergent and we need a~regularization of integral~(\ref{0.0}). Let $\chi\in
C_0 ^{\infty}(\mathbb{R}^{n})$, and $\chi(\boldsymbol{x})=1$ in a small
neighborhood of the origin. We set $\chi_{R}(\boldsymbol{x})=\chi(\boldsymbol{x}/R)$.

\begin{proposition}[\protect{\cite[Chapter~1]{Shubin}}]                                       
Let estimate \eqref{0.1} and conditions $(i)$--$(iii)$ hold. Then there exists a limit
\begin{gather}
\boldsymbol{F}=\lim_{R\rightarrow\infty}\int_{\mathbb{R}^{n}}\chi_{R}
(\boldsymbol{x})\boldsymbol{f}(\boldsymbol{x})e^{iS(\boldsymbol{x})}{\rm d}\boldsymbol{x} \label{0.2'}
\end{gather}
independent of the choose of the function~$\chi$.
\end{proposition}

\begin{proof}
We introduce the f\/irst-order dif\/ferential operator $L$
\begin{gather*}
Lu(\boldsymbol{x})=\big(1+\vert \nabla S(\boldsymbol{x})\vert^{2}\big)^{-1}
\big(I-i\nabla S(\boldsymbol{x})\cdot\nabla\big)u(\boldsymbol{x}),\qquad \boldsymbol{x}\in\mathbb{R}^{n}.
\end{gather*}
One can see that
\begin{gather}
Le^{iS(\boldsymbol{x},\boldsymbol{y})}=e^{iS(\boldsymbol{x},\boldsymbol{y})}. \label{0.4}
\end{gather}
Let $L^{\tau}$ be the transpose to $L$ dif\/ferential operator. Then taking into
the account~(\ref{0.4}) and applying the integration by parts we obtain
\begin{gather}
\boldsymbol{F}_{R}=\int_{\mathbb{R}^{n}}\chi_{R}(\boldsymbol{x})\boldsymbol{f}
(\boldsymbol{x})e^{iS(\boldsymbol{x})}{\rm d}\boldsymbol{x}
=\int_{\mathbb{R}^{n}}(L^{\tau})^{j}
\big(\chi_{R}(\boldsymbol{x})\boldsymbol{f}(\boldsymbol{x})\big)
e^{iS(\boldsymbol{x})}{\rm d}\boldsymbol{x}.\label{0.4'}
\end{gather}
Conditions $(i)$--$(iii)$ yield 
\begin{gather*}
\big\vert (L^{\tau})^{j}
\big(\chi_{R}(\boldsymbol{x})\boldsymbol{f}(\boldsymbol{x})\big)
\big\vert \leq C_{j} \langle
\boldsymbol{x} \rangle^{k-\rho j},
\end{gather*}
with the constant $C_{j}>0$ independent of $R>0$.
Let $j>\frac{k+n}{\rho}$, then the integral in the right side part of~(\ref{0.4'}) is absolutely
convergent, uniformly with respect to $R>0$, and we can go to the limit for
$R\rightarrow\infty$ in~(\ref{0.4'}). Hence the limit in~(\ref{0.2'}) exists,
independent of $\chi$, and
\begin{gather}
\boldsymbol{F} = \lim_{R\rightarrow\infty}\boldsymbol{F}_{R}=\int_{\mathbb{R}^{n}}\big((L^{{\tau}})
^{j}\boldsymbol{f}
(\boldsymbol{x})\big) e^{iS(\boldsymbol{x})}{\rm d}\boldsymbol{x}, \label{0.6}
\end{gather}
where $j>\frac{k+n}{\rho}$. The integrals def\/ined by formula~(\ref{0.6}) are
called \textit{oscillatory}.
\end{proof}

\subparagraph*{$3^{0}.$}
We consider an integral depending on a parameter $\lambda>0$
\[
\boldsymbol{I}_{\lambda}=\int_{\mathbb{R}^{n}}\boldsymbol{f}(\boldsymbol{x})e^{i\lambda
S(\boldsymbol{x})}{\rm d}\boldsymbol{x},
\]
where $\boldsymbol{f}$, $S$ satisfy condition (\ref{0.1}), $(i)$--$(iii)$. We say that
$\boldsymbol{x}_0 $ is a non-degenerate stationary point of the phase~$S$ if
\[
\nabla S(\boldsymbol{x}_0)=0,
\]
and
\[
\det S^{\prime\prime}(\boldsymbol{x}_0)\neq0,
\]
where $S^{\prime\prime}(\boldsymbol{x})=\left(\frac{\partial^{2}S(\boldsymbol{x})}{\partial x_{i}\partial
x_{j}}\right)_{i,j=1}^{n}$ is the Hess matrix of
the phase $S$ at the point $\boldsymbol{x}$.

\begin{proposition}[see for instance \cite{Fedoruk,BH}] \label{A1}Let there exist a
finite set $\left\{ \boldsymbol{x}_{1},\dots,\boldsymbol{x}_{N}\right\} $ of
non-degenerate stationary points of the phase $S$. Then
\begin{gather*}
\boldsymbol{I}_{\lambda}=\sum_{j=1}^{N}\boldsymbol{F}_{j}(\lambda), 
\end{gather*}
where
\begin{gather*}
\boldsymbol{F}_{j}(\lambda)=\left(\frac{2\pi}{\lambda}\right)^{\frac{n}{2}
}\frac{\exp\left(i\lambda S(\boldsymbol{x}_{j})+\frac{i\pi}{4}\operatorname{sgn} S^{\prime\prime
}(\boldsymbol{x}_{j})\right)}{\vert \det S^{\prime\prime}(\boldsymbol{x}_{j})\vert
^{1/2}}\boldsymbol{f}(\boldsymbol{x}_{j})\left(1+O\left(\frac{1}{\lambda}\right)\right),
\end{gather*}
and $\operatorname{sgn} S^{\prime\prime}(\boldsymbol{x}_{j})$ is the difference between the
number of positive and negative eigenvalues of the matrix $S^{\prime\prime}(\boldsymbol{x}_{j})$.
\end{proposition}

\section{Electromagnetic wave propagation in dispersive media}\label{section3}

The Maxwell equations in dispersive media are obtained by the replacing of the
electric and magnetic permittivity $\varepsilon$, $\mu$ by the operators
$\varepsilon(D_{t})$, $\mu(D_{t})$ where
\begin{gather*}
\varepsilon(D_{t})\boldsymbol{u}(t) =\frac{1}{2\pi}\int_{-\infty}^{\infty
}\varepsilon(\omega)\hat{\boldsymbol{u}}(\omega)e^{i\omega t}{\rm d}\omega,\qquad
\mu(D_{t})\boldsymbol{u}(t) =\frac{1}{2\pi}\int_{-\infty}^{\infty}\mu
(\omega)\hat{\boldsymbol{u}}(\omega)e^{i\omega t}{\rm d}\omega,
\end{gather*}
where the Fourier transform
\[
\hat{\boldsymbol{u}}(\omega)=\int_{-\infty}^{\infty}\boldsymbol{u}(t)e^{-i\omega t}{\rm d}t
\]
is understood in the sense of distributions. Let
\[
c(\omega)=\frac{1}{\sqrt{\varepsilon(\omega)\mu(\omega)}}
\]
be a phase velocity, $k(\omega)=\frac{\omega}{c(\omega)}$ be a wave number.

We suppose (see \cite[Chapter~IX]{LL8}) that:
\begin{enumerate}\itemsep=0pt
\item[$(i)$] the functions $\varepsilon(\omega),\mu(\omega)$ are limits in the sense of
the distributions of analytic bounded in the upper complex half-plane functions;
\item[$(ii)$] $k^{2}(\omega)$ has a f\/inite number $\omega_{1}<\cdots<\omega_{k}$ of simple
zeros on $\mathbb{R}$, and
$
\inf\limits_{\omega\in\mathbb{R}\backslash [ \omega_{1}-\epsilon,\omega
_{k}+\epsilon ] }k^{2}(\omega)>0
$
for small enough $\epsilon>0$;
\item[$(iii)$] the group velocity $v_{g}(\omega)=\frac{1}{k^{\prime}(\omega)}>0$ for
all $\omega\in\mathbb{R}\backslash [ \omega_{1}-\epsilon,\omega
_{k}+\epsilon ]$.
\end{enumerate}

The system of the Maxwell equations for dispersive medias is
\begin{gather*}
\boldsymbol{\nabla}\times \boldsymbol{H}
= \varepsilon(D_{t})\frac{\partial\boldsymbol{E}}{\partial t}+\boldsymbol{j},\nonumber 
\\
\boldsymbol{\nabla}\times \boldsymbol{E} =\mu(D_{t})\frac{\partial\boldsymbol{H}}{\partial t},\nonumber 
\\
\varepsilon(D_{t}) \boldsymbol{\nabla}\cdot \boldsymbol{E} = \rho,\nonumber 
\\
\boldsymbol{\nabla}\cdot \boldsymbol{H} =0,\nonumber 
\end{gather*}
with the continuity equation
\begin{gather*}
\boldsymbol{\nabla}\cdot \boldsymbol{j}+\frac{\partial\rho}{\partial t}=0. 
\end{gather*}
After the Fourier transform with respect to the time we obtain
\begin{gather}
\boldsymbol{\nabla}\times\hat{\boldsymbol{H}}(\omega) = i\varepsilon(\omega
)\omega\hat{\boldsymbol{E}}(\omega)+\hat{\boldsymbol{j}}(\omega),
\label{1.1'}\\
\boldsymbol{\nabla}\times \hat{\boldsymbol{E}}(\omega) =i\mu(\omega)\omega\hat{\boldsymbol{H}}(\omega),\label{1.2'}\\
\varepsilon(\omega)\boldsymbol{\nabla}\cdot \hat{\boldsymbol{E}}(\omega) =
\hat{\rho} (\omega),\nonumber 
\\
\boldsymbol{\nabla}\cdot \hat{\boldsymbol{H}}(\omega) =0, \nonumber   
\\
\boldsymbol{\nabla}\cdot \hat{\boldsymbol{j}}(\omega)-i\omega\hat{\rho
}(\omega)=0.\nonumber
\end{gather}
In the standard way system (\ref{1.1'}), (\ref{1.2'}) is reduced to the pair of
independent equations
\begin{gather}
\boldsymbol{\nabla}\times\boldsymbol{\nabla}\times\hat{\boldsymbol{E}}(\omega) - k^{2}(\omega
)\hat{\boldsymbol{E}}(\omega) = i\omega\mu(\omega)\hat{\boldsymbol{j}}(\omega), \label{1.7}
\\
\boldsymbol{\nabla}\times\boldsymbol{\nabla}\times\hat{\boldsymbol{H}}(\omega)-k^{2}(\omega)\hat{\boldsymbol{H}}(\omega)
=\boldsymbol{\nabla}\times\hat{\boldsymbol{j}}(\omega). \label{1.8}
\end{gather}
Unique solutions of equations (\ref{1.7}), (\ref{1.8}) in domain where
$k^{2}(\omega)>0$ is def\/ined by the limiting absorption principle. That is
\[
\hat{\boldsymbol{E}}(\omega)=\lim_{\epsilon\rightarrow+0}\hat{\boldsymbol{E}}_{\epsilon}(\omega),
\qquad \hat{\boldsymbol{H}}(\omega) = \lim_{\epsilon
\rightarrow+0}\hat{\boldsymbol{H}}_{\epsilon}(\omega),
\]
where $\hat{\boldsymbol{E}}_{\epsilon}(\omega)$, $\hat{\boldsymbol{H}}_{\epsilon}(\omega)$ are the unique bounded
solutions of the equations
\begin{gather*}
\boldsymbol{\nabla}\times
\boldsymbol{\nabla}\times\hat{\boldsymbol{E}}_{\epsilon}(\omega)-
\big(k^{2}(\omega)+i\epsilon\big)\hat{\boldsymbol{E}}_{\epsilon}(\omega) =
i\omega\mu(\omega)\hat{\boldsymbol{j}}(\omega),\\
\boldsymbol{\nabla}\times\boldsymbol{\nabla}\times\hat{\boldsymbol{H}}_{\epsilon}(\omega) -
\big(k^{2}(\omega)+i\epsilon\big)\hat{\boldsymbol{H}}_{\epsilon}(\omega)
=\boldsymbol{\nabla}\times\hat{\boldsymbol{j}}(\omega),
\end{gather*}
and $k(\omega)=i\sqrt{\vert k^{2}(\omega)\vert }$ is a purely
imaginary number in domains where $k^{2}(\omega)<0$. For these $\omega$
equations (\ref{1.7}), (\ref{1.8}) have unique decreasing solutions. By using
the relations
\[
\boldsymbol{\nabla}\times\boldsymbol{\nabla}\times\hat{\boldsymbol{E}}(\omega)=-\boldsymbol{\nabla}
^{2}\hat{\boldsymbol{E}}(\omega) + \boldsymbol{\nabla}(\boldsymbol{\nabla}\cdot\hat{\boldsymbol{E}}(\omega))
\]
and
\begin{gather*}
\boldsymbol{\nabla} \cdot\hat{\boldsymbol{E}}(\omega) = \varepsilon^{-1}(\omega)\hat
{\rho}(\omega), 
\end{gather*}
we reduce (\ref{1.7}) to the equation
\begin{gather}
\boldsymbol{\nabla}^{2}\hat{\boldsymbol{E}}(\omega) + k^{2}(\omega
)\hat{\boldsymbol{E}}(\omega) = \varepsilon^{-1}(\omega)\boldsymbol{\nabla}
\hat{\rho}(\omega)-i\omega\mu(\omega)\hat{\boldsymbol{j}}(\omega
)\nonumber\\
\hphantom{\boldsymbol{\nabla}^{2}\hat{\boldsymbol{E}}(\omega) + k^{2}(\omega)\hat{\boldsymbol{E}}(\omega)}{}
=-i\omega\mu(\omega)\left(\hat{\boldsymbol{j}}(\omega) +
\frac{1}{k^{2}(\omega)}\boldsymbol{\nabla}(\boldsymbol{\nabla}\cdot\hat{\boldsymbol{j}})(\omega)\right)
=\boldsymbol{F}_{\omega}.\label{1.9'}
\end{gather}
The similar way we obtain from equation (\ref{1.8}) that
\begin{gather}
\boldsymbol{\nabla}^{2}\hat{\boldsymbol{H}}(\omega) + k^{2}(\omega
)\hat{\boldsymbol{H}}(\omega)=-\boldsymbol{\nabla}\times\hat{\boldsymbol{j}}(\omega
)=\boldsymbol{\Phi}_{\omega}. \label{1.11}
\end{gather}
Equations (\ref{1.9'}) and (\ref{1.11}) are independent and can be used for
the def\/inition of $\hat{\boldsymbol{E}}$ and $\hat{\boldsymbol{H}}$.

Let
$
g_{\omega}(\boldsymbol{x})=\frac{e^{ik(\omega)\vert \boldsymbol{x}\vert }
}{4\pi\vert \boldsymbol{x}\vert }
$ 
be the fundamental solution of the scalar Helmholtz equation
\[
\Delta g_{\omega}(\boldsymbol{x})+k^{2}(\omega)g_{\omega}(\boldsymbol{x})=- \delta(\boldsymbol{x}),\qquad
\boldsymbol{x}\in\mathbb{R}^{3},
\]
satisfying the limiting absorption principle. Hence the solutions of equations
(\ref{1.9'}), (\ref{1.11}) are given as
\begin{gather}\label{1.11'}
\boldsymbol{E}(t, x) = \frac{1}{2\pi}\int_{-\infty}^{\infty
}e^{i\omega t}(g_{\omega}\ast\boldsymbol{F}_{\omega})(\boldsymbol{x}){\rm d}\omega
,\qquad
\boldsymbol{H}(t,\boldsymbol{x}) = \frac{1}{2\pi}\int_{-\infty}^{\infty
}e^{i\omega t}(g_{\omega}\ast\boldsymbol{\Phi}_{\omega})(\boldsymbol{x}){\rm d}\omega.
\end{gather}
where the convolution is understood in the sense of the distributions
\[
(g\ast\boldsymbol{\Psi})(\boldsymbol{x})=\int_{\mathbb{R}^{3}}
\frac{e^{ik(\omega)\vert \boldsymbol{x}-\boldsymbol{y}\vert }}{4\pi\vert\boldsymbol{x}-\boldsymbol{y}\vert }
\boldsymbol{\Psi}(y){\rm d}\boldsymbol{y}.  
\]

Let
\begin{gather}
\boldsymbol{j}(t,\boldsymbol{x})=A(t)\boldsymbol{v}(t)\delta(\boldsymbol{x}-\boldsymbol{x}
_0 (t)), \label{1.12}
\end{gather}
where $\delta$ is the standard $\delta$-function, $\boldsymbol{v}(t)=\dot{\boldsymbol{x}}_0 (t)$
is a velocity of a source, $A(t)$ is an amplitude of the source.
Then (\ref{1.11'}) implies that
\begin{gather}
\boldsymbol{H}(t,\boldsymbol{x})=\frac{1}{8\pi^{2}}\int_{\mathbb{R}^{2}}
\nabla_{\boldsymbol{x}}\times\left(\frac{e^{i(k(\omega)\left\vert \boldsymbol{x}-\boldsymbol{x}
_0 (\tau)\right\vert -\omega(t-\tau))}}{\left\vert \boldsymbol{x}-\boldsymbol{x}_0
(\tau)\right\vert }\boldsymbol{v}(\tau)\right) A(\tau){\rm d}\omega{\rm d}\tau, \label{1.13}
\end{gather}
and
\begin{gather}
\boldsymbol{E}(t,\boldsymbol{x})=\frac{1}{8\pi^{2}i}\int_{\mathbb{R}^{2}}
A(\tau)\omega\mu(\omega)\big(I+k^{-2}(\omega)\nabla_{\boldsymbol{x}}\nabla
_{\boldsymbol{x}}\cdot\big)\frac{e^{i(k(\omega)\left\vert \boldsymbol{x}-\boldsymbol{x}_0      
(\tau)\right\vert -\omega(t-\tau))}}{\left\vert \boldsymbol{x}-\boldsymbol{x}_0
(\tau)\right\vert }\boldsymbol{v}(\tau){\rm d}\omega{\rm d}\tau.\!\!\! \label{1.13'}
\end{gather}

Applying the analyticity of the integrand with respect to $\omega$ in the
upper half-plane $\mathbb{C}_{+}$ we deform the line of integration
$(-\infty,\infty)$ with respect to $\omega$ into the contour $\Gamma
=(-\infty,\omega_{1}-\varepsilon)\cup\Gamma^{\prime}\cup(\omega_{k}
+\varepsilon,+\infty)$, where $\Gamma^{\prime}$ is located in the upper
complex half-plane and bypasses from above all singularities $\left\{
\omega_{1},\dots,\omega_{k}\right\} $ of the integrand on the real line.

The phase of the double integrals is
\[
S(\omega,\tau)=k(\omega)\left\vert \boldsymbol{x}-\boldsymbol{x}_0 (\tau)\right\vert
+\omega\tau,
\]
and
\begin{align*}
\frac{\partial S(\omega,\tau)}{\partial\omega}=\frac{\left\vert
\boldsymbol{x}-\boldsymbol{x}_0 (\tau)\right\vert }{v_{g}(\omega)}+\tau,
\qquad
\frac{\partial S(\omega,\tau)}{\partial\tau}=-k(\omega)v(\tau,\boldsymbol{x})+\omega,
\end{align*}
where
\[
v(\tau,\boldsymbol{x})=\frac{\dot{\boldsymbol{x}}_0 (\tau)\cdot(\boldsymbol{x}-\boldsymbol{x}_0
(\tau))}{\left\vert \boldsymbol{x}-\boldsymbol{x}_0 (\tau)\right\vert }
\]
is the projection of the speed $\dot{\boldsymbol{x}}_0 (\tau)$ on the vector
$\boldsymbol{x}-\boldsymbol{x}_0 (\tau)$. We suppose that there exists 
large enough $R>0$ such
that
\begin{gather*}
\inf_{\omega^{2}+\tau^{2}\geq R^{2}}\left\vert \frac{v(\tau,\boldsymbol{x})}
{v_{g}(\omega)}-1\right\vert>0,
\qquad
\inf_{\omega^{2}+\tau^{2}\geq R^{2}}\left\vert \frac{v(\tau,\boldsymbol{x})
}{c(\omega)}-1\right\vert>0.
\end{gather*}
Then the phase $S$ in integrals (\ref{1.13}), (\ref{1.13'}) satisf\/ies the
estimate
\[
\left\vert \nabla S(\omega,\tau)\right\vert \geq C(\left\vert \omega
\right\vert +\left\vert \tau\right\vert)
\]
in the domain $\left\{ (\omega,\tau)\in\mathbb{R}^{2}:\omega^{2}+\tau^{2}\geq
R^{2}\right\} $ where $C=C(R)>0$ for $R>0$ large enough. Hence integrals
(\ref{1.13}), (\ref{1.13'}) exist as oscillatory.

\subsection{Asymptotic analysis of the f\/ields of moving sources}\label{section3.1}

Now we introduce a dimensionless parameter $\lambda>0$ as
\[
\lambda=\left(\inf_{t\in\mathbb{R}}\left\vert \boldsymbol{x}-\boldsymbol{x}
_0 (t)\right\vert \right) \frac{\Omega}{c_0 }>0,
\]
where $c_0 $ is the light speed in the vacuum, $\Omega>0$ is a characteristic
frequency of the problem. In what follows we suppose that $\lambda$ is a large
parameter, that frequently is a ratio of distance between the moving source
and the receiver to the f\/ield wavelength. Such a distance is much more then
$\left(\frac{\Omega}{c_0 }\right)^{-1}$ for all the time. Since
$a^{\prime}(t)=\frac{1}{\lambda}\tilde{a}^{\prime}(t/\lambda)=O(\frac
{1}{\lambda})$ (see equation~\eqref{s1'}) the $1/\lambda$ in general characterizes the
slowness variations of the amplitude $a$ due to slowness variations of
dif\/ferent parameters of a problem, e.g.\ \textit{charge trajectory, material
dispersion, etc}.

In what follows we will suppose that
\begin{gather}
A(t)=a(t)e^{-i\omega_0 t}, \label{s1}
\end{gather}
where
\begin{gather}
a(t)=\tilde{a}(t/\lambda), \label{s1'}
\end{gather}
$\tilde{a}\in C_{b}^{\infty}(\mathbb{R)}$, $\omega_0 >0$ is an eigenfrequency
(a carrier frequency) of the source,
\begin{gather}
\boldsymbol{x}_0 (t)=\lambda\boldsymbol{X}_0(t/\lambda),\qquad t\in\mathbb{R},
\label{s2}
\end{gather}
$\lambda>0$ is a large dimensionless parameter characterizing the slowness of
variations of the amplitude $a$, and the velocity
\[
\dot{\boldsymbol{x}}_0 (t)=\dot{\boldsymbol{X}}_0(t/\lambda),\qquad t\in
\mathbb{R},
\]
where the vector-function $\dot{\boldsymbol{X}}_0  (t)\in C_{b}^{\infty
}(\mathbb{R)\otimes C}^{3}$.

To reduce integrals (\ref{1.13}), (\ref{1.13'}) to a form containing the large
parameter $\lambda>0$ we use the scale change of variables
\begin{gather*}
\boldsymbol{x}=\lambda\boldsymbol{X},\qquad t=\lambda T,\qquad \tau=\lambda\iota. 
\end{gather*}
and obtain following representations for magnetic and electric f\/ields
\begin{gather}
\bar{\boldsymbol{H}}_{\lambda}(T,\boldsymbol{X})=\frac{1}{8\pi^{2}\lambda}
\int_{\Gamma\times\mathbb{R}}\tilde{a}(\iota)\nabla_{\boldsymbol{X}}
\times\left(\frac{e^{i\lambda\bar{S}(T,\boldsymbol{X},\omega,\iota)}
}{\left\vert \boldsymbol{X} - \boldsymbol{X}_0 (\iota)\right\vert }\boldsymbol{V}(\iota)\right)
{\rm d}\omega {\rm d}\iota,\label{1.16}
\\
\bar{\boldsymbol{E}}_{\lambda}(t,\boldsymbol{x})=\frac{1}{8\pi^{2}i}\int_{\Gamma\times\mathbb{R}}\tilde{a}(\iota)\omega\mu(\omega)\Bigg(I+\frac
{1}{\lambda^{2}k^{2}(\omega)}\nabla_{\boldsymbol{x}}\nabla_{\boldsymbol{x}}\cdot
\Bigg)\frac{e^{i\lambda\bar{S}(T,\boldsymbol{X},\omega,\iota)}}{\left\vert
\boldsymbol{X} - \boldsymbol{X}_0(\iota)\right\vert}\boldsymbol{V}(\iota)
{\rm d}\omega{\rm d}\iota.\label{1.16'}
\end{gather}
The phase $\bar{S}$ in integrals (\ref{1.16}), (\ref{1.16'})  at $\lambda \rightarrow +\infty$ is
\begin{gather*}
\bar{S}(T,\boldsymbol{X},\omega,\iota)=k(\omega)\left\vert \boldsymbol{X} - \boldsymbol{X}
_0 (\iota)\right\vert -\omega(T-\iota)-\omega_0 \iota. 
\end{gather*}
Note that contributions in the main term of asymptotics of integrals
(\ref{1.16}), (\ref{1.16'}) are given by the stationary points of the phase
$\bar{S}(T,\boldsymbol{X},\omega,\iota)$ located in the domain
$\mathbb{R}\backslash\left[ \omega_{1}-\epsilon,\omega_{k}+\epsilon\right]
$. The stationary points of $\bar{S}(T,\boldsymbol{X},\omega,\iota)$ with
respect to $(\omega,\iota)$ for f\/ixed $(T,\boldsymbol{X})$ are solutions
of the system
\begin{gather}
\frac{\partial\bar{S}(T,\boldsymbol{X},\omega,\iota)}{\partial\omega}
=\frac{\left\vert \boldsymbol{X} - \boldsymbol{X}_0 (\iota)\right\vert }{v_{g}(\omega)}
-(T-\iota)=0,\nonumber\\
\frac{\partial\bar{S}(T,\boldsymbol{X},\omega,\iota)}{\partial\iota}
=-k(\omega)V(\boldsymbol{X},\tau)+(\omega-\omega_0)=0,\label{1.18}
\end{gather}
where
\[
V(\boldsymbol{X},\tau)
=\frac{\boldsymbol{X} - \boldsymbol{X}_0 (\iota)}{\left\vert \boldsymbol{X} -
 \boldsymbol{X}_0(\iota)\right\vert }\cdot\boldsymbol{V}(\iota)
\]
is the value of the projection of $\boldsymbol{V}(\iota)$ on the vector
$\boldsymbol{X} - \boldsymbol{X}_0 (\iota)$.

Let
$
\omega_{s}=\omega_{s}(T,\boldsymbol{X})$, $\iota_{s}=\iota_{s}(T,\boldsymbol{X})
$ 
be a non-degenerate stationary point of the phase $\bar{S}$. It~means that
$(\omega_{s},\iota_{s})$ is a solution of system (\ref{1.18}) and
\begin{gather*}
\det\bar{S}^{\prime\prime}(T,\boldsymbol{X},\omega_{s},\iota_{s})\neq0,
\end{gather*}
where
\begin{gather*}
\bar{S}^{\prime\prime}(T,\boldsymbol{X},\omega,\iota)= \begin{pmatrix}
k^{\prime\prime}(\omega)\left\vert \boldsymbol{X} - \boldsymbol{X}_0 (\iota)\right\vert &
1-\dfrac{V(\boldsymbol{X},\tau)}{v_{g}(\omega)}\vspace{1mm}\\
1-\dfrac{V(\boldsymbol{X},\tau)}{v_{g}(\omega)} & -k(\omega)\dfrac{\partial
V(\boldsymbol{X},\iota)}{\partial\iota}
\end{pmatrix}
\end{gather*}
is the Hess matrix of the phase $\bar{S}$. We denote by $\operatorname{sgn}\bar{S}
^{\prime\prime}(T,\boldsymbol{X},\omega_{s},\iota_{s})$ the dif\/ference
between the number of positive and negative eigenvalues of the matrix $\bar
{S}^{\prime\prime}(T,\boldsymbol{X},\omega_{s},\iota_{s})$.

The contribution of the stationary point $(\omega_{s},\iota_{s})$ in
the asymptotics of $\bar{\boldsymbol{H}}_{\lambda}(T,\boldsymbol{X})$,
$\bar{\boldsymbol{E}}_{\lambda}(T,\boldsymbol{X})$ is given by the formulae (see Proposition~\ref{A1})
\begin{gather}
\bar{\boldsymbol{H}}_{\lambda,s}(T,\boldsymbol{X})
 =\frac{1}{4\pi\lambda^{2}}
\nabla_{\boldsymbol{X}}\times\left(\frac{e^{i\lambda\bar{S}(T,\boldsymbol{X},
\omega_{s},\iota_{s})}}{\left\vert \boldsymbol{X} - \boldsymbol{X}_0 (\iota
_{s})\right\vert }\boldsymbol{V}(\iota_{s})\right) \frac{e^{\frac{i\pi}{4}
\operatorname{sgn}\bar{S}^{\prime\prime}(T,\boldsymbol{X},\omega_{s},\iota_{s})}
}{\left\vert \det\bar{S}^{\prime\prime}(T,\boldsymbol{X},\omega_{s},
\iota_{s})\right\vert^{1/2}}\tilde{a}(\iota_{s})\nonumber\\
\hphantom{\bar{\boldsymbol{H}}_{\lambda,s}(T,\boldsymbol{X})=}{}
\times\left(1+O\left(\frac{1}{\lambda}\right)\right)\label{1.20'}
\end{gather}
and
\begin{gather}
\bar{\boldsymbol{E}}_{\lambda,s}(t,\boldsymbol{x})   =\frac{\tilde{a}(\iota_{s})}{4\pi\lambda i}
\left[\omega_{s}\mu(\omega_{s})I+\frac{1}{\lambda^{2}k^{2}
(\omega_{s})}\nabla_{\boldsymbol{x}}\nabla_{\boldsymbol{x}}\cdot\frac{e^{i\lambda
\bar{S}(T,\boldsymbol{X},\omega_{s},\iota_{s})}}{\left\vert \boldsymbol{X} - \boldsymbol{X}_0
(\iota_{s})\right\vert }\boldsymbol{V}(\iota_{s})\right]\nonumber\\
\hphantom{\bar{\boldsymbol{E}}_{\lambda,s}(t,\boldsymbol{x})=}{}
 \times\frac{e^{\frac{i\pi}{4}\operatorname{sgn}\bar{S}^{\prime\prime}(T,\boldsymbol{X},
\omega_{s},\iota_{s})}}{\left\vert \det\bar{S}^{\prime\prime
}(T,\boldsymbol{X},\omega_{s},\iota_{s})\right\vert^{1/2}}\left(1+O\left(\frac{1}{\lambda}\right)\right).\label{1.21}
\end{gather}

If the phase $\bar{S}$ has a f\/inite set of stationary points the main term of
the asymptotics of the electromagnetic f\/ield is the sum of contributions of
the every stationary point.

The expressions (\ref{1.20'}) and (\ref{1.21}) can be simplif\/ied if we are
restricted by the terms of the order $O\left(\frac{1}{\lambda}\right)$
\begin{gather}
\bar{\boldsymbol{H}}_{\lambda,s}(T,\boldsymbol{X})\nonumber\\
{} \sim\frac{ik(\omega_{s})
 \tilde
{a}(\iota_{s})}{4\pi\lambda}\frac{e^{i\lambda\bar{S}(T,\boldsymbol{X},\omega
_{s},\iota_{s})+\frac{i\pi}{4}\operatorname{sgn}\bar{S}^{\prime\prime}
(T,\boldsymbol{X},\omega_{s},\iota_{s})}(\nabla_{\boldsymbol{X}} \times \boldsymbol{V}(\iota_{s}))
\left\vert\boldsymbol{X} - \boldsymbol{X}_0 (\iota_{s})\right\vert
}{\left\vert \det\bar{S}^{\prime\prime}(T,\boldsymbol{X},\omega_{s},
\iota_{s})\right\vert^{1/2}\left\vert\boldsymbol{X} - \boldsymbol{X}_0 (\iota_{s})\right\vert}\label{1.21'}
\end{gather}
and
\begin{gather}
\bar{\boldsymbol{E}}_{\lambda,s}(t,\boldsymbol{x})   \sim\frac{\tilde{a}(\iota_{s})}{4\pi\lambda i}
\left[\omega_{s}\mu(\omega_{s})\boldsymbol{V}(\iota_{s})
-\big(\nabla_{\boldsymbol{x}}\nabla_{\boldsymbol{x}}\cdot\boldsymbol{V}(\iota_{s})\big)
\left\vert \boldsymbol{X} - \boldsymbol{X}_0 (\iota_{s})\right\vert \right]\nonumber
\\
\hphantom{\bar{\boldsymbol{E}}_{\lambda,s}(t,\boldsymbol{x})\sim}{}
 \times\frac{e^{i\lambda\bar{S}(T,\boldsymbol{X},\omega_{s},\iota
_{s})+\frac{i\pi}{4}\operatorname{sgn}\bar{S}^{\prime\prime}(T,\boldsymbol{X},\omega
_{s},\iota_{s})}}{\left\vert \boldsymbol{X} - \boldsymbol{X}_0 (\iota_{s})\right\vert
\left\vert \det\bar{S}^{\prime\prime}(T,\boldsymbol{X},\omega_{s},
\iota_{s})\right\vert^{1/2}}. \label{1.21''}
\end{gather}
Coming back to the variables $(t,\boldsymbol{x})$ we obtain the following
asymptotic formulae
\begin{gather}
\boldsymbol{H}_{s}(t,\boldsymbol{x})\sim\frac{1}{4\pi}\nabla_{\boldsymbol{x}}
\times\left(\frac{e^{iS(t,\boldsymbol{x},\omega_{s},\tau_{s})}}{\left\vert
\boldsymbol{x}-\boldsymbol{x}_0 (\tau_{s})\right\vert }\boldsymbol{v}(\tau_{s})\right)
\frac{a(\tau_{s})e^{\frac{i\pi}{4}\operatorname{sgn}S^{\prime\prime}(t,\boldsymbol{x},\omega_{s},
\tau_{s})}}{\left\vert \det S^{\prime\prime}(t,\boldsymbol{x},
\omega_{s},\tau_{s})\right\vert^{1/2}}, \label{h2}
\\
\boldsymbol{E}_{s}(t,\boldsymbol{x})
\sim\frac{1}{4\pi i}a(\tau
_{s})\omega_{s}\mu(\omega_{s})\left(I+\frac{1}{k^{2}(\omega_{s})}\nabla
_{\boldsymbol{x}}\nabla_{\boldsymbol{x}}\cdot\right)\frac{e^{iS(t,\boldsymbol{x},\omega
_{s},\tau_{s})}}{\left\vert \boldsymbol{x}-\boldsymbol{x}_0 (\tau_{s})\right\vert
}\boldsymbol{v}(\tau_{s}) \nonumber\\
\phantom{\boldsymbol{E}_{s}(t,\boldsymbol{x})\sim}
\times\frac{e^{\frac{i\pi}{4}\operatorname{sgn}S^{\prime\prime}(t,\boldsymbol{x},\omega
_{s},\tau_{s})}}{\left\vert \det S^{\prime\prime}(t,\boldsymbol{x},
\omega_{s},\tau_{s})\right\vert^{1/2}},\label{h3}
\end{gather}
where
\begin{gather*}
t=\frac{T}{\lambda},
\qquad
\left\vert \boldsymbol{x}-\boldsymbol{x}_0 (t)\right\vert =\frac
{\left\vert \boldsymbol{X} - \boldsymbol{X}_0 (T)\right\vert }{\lambda},
\qquad
\lambda\rightarrow\infty. 
\end{gather*}
In formulae (\ref{h2}), (\ref{h3}) the phase is
$
S(t,\boldsymbol{x},\omega,\tau)=k(\omega) \vert \boldsymbol{x}-\boldsymbol{x}_0
(\tau) \vert -\omega(t-\tau)-\omega_0 \tau,
$
and the stationary points $\big(\omega_{s}=\omega_{s}(t,\boldsymbol{x}),\tau_{s}
=\tau_{s}(t,\boldsymbol{x})\big)$ are solutions of the system
\begin{gather}
\frac{\partial S(t,\boldsymbol{x},\omega,\tau)}{\partial\omega}
=\frac{\left\vert \boldsymbol{x}-\boldsymbol{x}_0(\tau)\right\vert}{v_{g}(\omega)}
-(t-\tau)=0,\nonumber\\
\frac{\partial S(t,\boldsymbol{x},\omega,\tau)}{\partial\tau}
=-k(\omega)v(\boldsymbol{x},\tau)+(\omega-\omega_0)=0,\label{1.25}
\end{gather}
and
\[
S^{\prime\prime}(\boldsymbol{x},t,\omega,\tau)= \begin{pmatrix}
k^{\prime\prime}(\omega)\left\vert \boldsymbol{x}-\boldsymbol{x}_0 (\tau)\right\vert &
1-\dfrac{v(\boldsymbol{x},\tau)}{v_{g}(\omega)}\vspace{1mm}\\
1-\dfrac{v(\boldsymbol{x},\tau)}{v_{g}(\omega)} & -k(\omega)\dfrac{\partial
v(\boldsymbol{x},\tau)}{\partial\tau}
\end{pmatrix}.
\]
Note that under conditions
\begin{gather*}
\sup_{(\omega,\tau)\in\mathbb{R}^{2}}\left(\frac{\left\vert \boldsymbol{v}(
\tau)\right\vert }{\left\vert v_{g}(\omega)\right\vert }+\frac{\Omega}{T_0
}\left\vert k^{\prime\prime}(\omega)\right\vert \left\vert \boldsymbol{x}-\boldsymbol{x}
_0 (\tau)\right\vert \right)<1,\nonumber\\
\sup_{(\omega,\tau)\in\mathbb{R}^{2}}\left(\frac{T_0 }{\omega^{0}}
k(\omega)\left\vert \frac{\partial v(\boldsymbol{x},\tau)}{\partial\tau
}\right\vert +\frac{\left\vert \boldsymbol{v}(\tau)\right\vert }{\left\vert
v_{g}(\omega)\right\vert }\right)<1, 
\end{gather*}
where $(T_0,\Omega)$ are the scale time and frequency, system (\ref{1.25})
has an unique solution which can be obtained by the method of successive approximations.

Coming to the variables $(t,\boldsymbol{x})$ in formulae (\ref{1.21'}),
(\ref{1.21''}) we simplify formulae (\ref{h2}), (\ref{h3})
\begin{gather}
\boldsymbol{H}_{s}(t,\boldsymbol{x})\sim\frac{ik(\omega_{s})a(\tau_{s})}{4\pi}
\frac{e^{i\lambda S(t,\boldsymbol{x},\omega_{s},\tau_{s})+\frac{i\pi}
{4}\operatorname{sgn}S^{\prime\prime}(t,\boldsymbol{x},\omega_{s},\tau_{s})}
\big(\nabla_{\boldsymbol{x}}\times\boldsymbol{v}(\tau_{s})\big)
\big(\left\vert \boldsymbol{x}-\boldsymbol{x}_0
(\tau_{s})\right\vert\big)}{\left\vert \det S^{\prime\prime}(t,\boldsymbol{x},
\omega_{s},\iota_{s})\right\vert^{1/2}\left\vert \boldsymbol{x}-\boldsymbol{x}_0
(\tau_{s})\right\vert}, \label{h2'}
\\
\boldsymbol{E}_{s}(t,\boldsymbol{x})\sim\frac{a(\iota_{s})}{4\pi i}
\left[
\omega_{s}\mu(\omega_{s})\boldsymbol{v}(\tau_{s})-\big(\nabla_{\boldsymbol{x}}
\nabla_{\boldsymbol{x}}\cdot\boldsymbol{v}(\tau_{s})\big)
\big(\left\vert \boldsymbol{x}-\boldsymbol{x}_0
(\tau_{s})\right\vert\big)
\right]\nonumber\\
\phantom{\boldsymbol{E}_{s}(t,\boldsymbol{x})\sim}
\times\frac{e^{i\lambda S(t,\boldsymbol{x},\omega_{s},\tau_{s}
)+\frac{i\pi}{4}\operatorname{sgn}S^{\prime\prime}(t,\boldsymbol{x},\omega_{s},\tau
_{s})}}{\left\vert \boldsymbol{x}-\boldsymbol{x}_0 (\tau_{s})\right\vert \left\vert \det
S^{\prime\prime}(t,\boldsymbol{x},\omega_{s},\tau_{s})\right\vert^{1/2}}.\label{h3'}
\end{gather}

\begin{example}
Let $\boldsymbol{x}_0  (\tau)=(0,v\tau,H)$. Then $\boldsymbol{x}-\boldsymbol{x}
_0  (\tau)=(x_{1},x_{2}-v\tau,x_{3}-H)$,
\begin{gather*}
 \vert \boldsymbol{x}-\boldsymbol{x}_0 (\tau) \vert =\left(x_{1}^{2}+(x_{2}
-v\tau)^{2}+(x_{3} -H)^{2}\right)^{1/2}, 
\end{gather*}
$\boldsymbol{v}=(0,v,0)$. The system for the stationary phase point
$\big(\omega_{s}(t,\boldsymbol{x}),\tau_{s}(t,\boldsymbol{x})\big)$ is
\begin{gather*}
\frac{\left(x_{1}^{2}+(x_{2}-v\tau)^{2}+(x_{3}{}-H)^{2}\right)^{1/2}
}{v_{g}(\omega)}-(t-\tau)=0,\nonumber\\
-k(\omega)\frac{v(x_{2}-v\tau)}{\left(x_{1}^{2}+(x_{2}-v\tau)^{2}+x_{3}
^{2}\right)^{1/2}}+(\omega-\omega_0)=0.
\end{gather*}
For applying formulae (\ref{h2'}), (\ref{h3'}) we have to use
\begin{gather*}
(\nabla_{\boldsymbol{x}}\times\boldsymbol{v})\left\vert \boldsymbol{x}-\boldsymbol{x}_0
(\tau)\right\vert   =\left(-v\frac{\partial\left\vert \boldsymbol{x}-\boldsymbol{x}
_0 (\tau)\right\vert }{\partial x_{3}},0,v\frac{\partial\left\vert
\boldsymbol{x}-\boldsymbol{x}_0 (\tau)\right\vert }{\partial x_{1}}\right) \\
\hphantom{(\nabla_{\boldsymbol{x}}\times\boldsymbol{v})\left\vert \boldsymbol{x}-\boldsymbol{x}_0
(\tau)\right\vert}{}
 =\left(-v\frac{x_{3}-H}{\left\vert \boldsymbol{x}-\boldsymbol{x}_0 (\tau)\right\vert
},0,v\frac{x_{1}}{\left\vert \boldsymbol{x}-\boldsymbol{x}_0 (\tau)\right\vert }\right),
\end{gather*}
and
\[
(\nabla_{\boldsymbol{x}}\nabla_{\boldsymbol{x}}\cdot\boldsymbol{v})
\big(\left\vert\boldsymbol{x}-\boldsymbol{x}_0 (\tau)\right\vert\big)
=v\left(
\frac{x_{1}(x_{2}-v\tau)}{\left\vert \boldsymbol{x}-\boldsymbol{x}_0(\tau)\right\vert^{2}},
\frac{x_{1}^{2}+(x_{3}{}-H)^{2}}{\left\vert \boldsymbol{x}-\boldsymbol{x}_0 (\tau)\right\vert^{3}},
\frac{(x_{3}-H)(x_{2}-v\tau)}{\left\vert \boldsymbol{x}-\boldsymbol{x}_0 (\tau)\right\vert^{2}}
\right).
\]
\end{example}

\subsection{Doppler ef\/fect and retarded time}\label{section3.2}

Note that for f\/ix point $\boldsymbol{x}$ formulae (\ref{h2}), (\ref{h3}) can be
written of the form
\begin{gather*}
\boldsymbol{W}(t)=\boldsymbol{A}(t)e^{iF(t)}, 
\end{gather*}
where $\boldsymbol{A}(t)$ is a bounded vector-function, $F$ is a real-valued
function such that $\lim\limits_{t\rightarrow\infty}F(t)=\infty$. According to the
signal processing theory (see for instance \cite{Cohen}) $F(t)$ is a phase of
the wave process $\boldsymbol{W}(t)$, and the instantaneous frequency $\omega
_{\rm in}(t)$ of the wave process $\boldsymbol{W}(t)$ is def\/ined as
$
\omega_{\rm in}(t)=-F^{\prime}(t)$.
In our case
\begin{gather*}
F(t) =S\big(t,\boldsymbol{x},\omega_{s}(t,\boldsymbol{x}),\tau_{s}
(t,\boldsymbol{x})\big)\nonumber\\
\hphantom{F(t)}{} =k\big(\omega_{s}(t,\boldsymbol{x})\big)
\left\vert \boldsymbol{x}-\boldsymbol{x}_0\big(\tau_{s}(t,\boldsymbol{x})\big)\right\vert
-\omega_{s}(t,\boldsymbol{x})\big(t-\tau_{s}(t,\boldsymbol{x})\big)-\omega_0 \tau_{s}(t,\boldsymbol{x}),
\end{gather*}
where $\left(\omega_{s}(t,\boldsymbol{x}),\tau_{s}(t,\boldsymbol{x}
)\right)$ is a stationary point of the phase $S$. Dif\/ferentiating of $F$ as
a composed function we obtain
\begin{gather*}
-F^{\prime}(t)= -\frac{\partial S\big(t,\boldsymbol{x},\omega_{s}(t,\boldsymbol{x}),\tau_{s}(t,\boldsymbol{x})\big)}
{\partial t}-\frac{\partial S\big(t,\boldsymbol{x},\omega_{s}(t,\boldsymbol{x}),\tau_{s}(t,\boldsymbol{x})\big)}
{\partial\omega}\frac{\partial\omega_{s}(t,\boldsymbol{x})}{\partial t}\\
\hphantom{-F^{\prime}(t)=}{}
-\frac{\partial S\big(t,\boldsymbol{x},\omega_{s}(t,\boldsymbol{x}),\tau
_{s}(t,\boldsymbol{x})\big)}{\partial\tau}\frac{\partial\tau_{s}(t,\boldsymbol{x}
)}{\partial t}.
\end{gather*}
Taking into account that $\left(\omega_{s}(t,\boldsymbol{x}),\tau
_{s}(t,\boldsymbol{x})\right) $ is the stationary point of $S$, we obtain that
\[
\omega_{\rm in}(t)=\omega_{s}(t,\boldsymbol{x}).
\]
It implies that the instantaneous frequency $\omega_{\rm in}(t)$ of the wave
processes $\boldsymbol{H}(t,\boldsymbol{x})$, $\boldsymbol{E}(t,\boldsymbol{x})$ for f\/ixed~$\boldsymbol{x}$
coincides with $\omega_{s}(t,\boldsymbol{x})$. Hence the instantaneous Doppler
ef\/fect is
\begin{gather*}
\omega_{s}(t,\boldsymbol{x})-\omega_0 =k(\omega_{s}(t,\boldsymbol{x}))v(\boldsymbol{x},
\tau_{s}(t,\boldsymbol{x})). 
\end{gather*}
Considering the case $k(\omega_{s}(t,\boldsymbol{x}))>0$ we obtain the usual
Doppler ef\/fect
\begin{gather*}
v(\boldsymbol{x},\tau_{s}(t,\boldsymbol{x}))>0\ \Longrightarrow\ \omega_{s}(t,\boldsymbol{x}
)>\omega_0
\qquad \mbox{and}
\qquad
v(\boldsymbol{x},\tau_{s}(t,\boldsymbol{x}))<0\ \Longrightarrow\ \omega_{s}(t,\boldsymbol{x}
)<\omega_0 .
\end{gather*}

In the case $k(\omega_{s}(t,\boldsymbol{x}))<0$ (metamaterials) we obtain the
inverse Doppler ef\/fect
\begin{gather*}
v(\boldsymbol{x},\tau_{s}(t,\boldsymbol{x}))<0\ \Longrightarrow\ \omega_{s}(t,\boldsymbol{x}
)<\omega_0,
\qquad \mbox{and}
\qquad
v(\boldsymbol{x},\tau_{s}(t,\boldsymbol{x}))>0\ \Longrightarrow\ \omega_{s}(t,\boldsymbol{x}
)>\omega_0 .
\end{gather*}

It follows from formulae (\ref{h2'}), (\ref{h3'}) that $\tau_{s}
(t,\boldsymbol{x})$ is the excitation time of the signal arriving to the
receiver located at the point $\boldsymbol{x}$ at the time $t$. Hence the mode
Doppler ef\/fect for the time (the retarded time) is
\[
t-\tau_{s}(t,\boldsymbol{x})=\frac{\left\vert \boldsymbol{x}-\boldsymbol{x}_0 (\tau_{s}
(t,\boldsymbol{x}))\right\vert }{v_{g}(\omega_{s}(t,\boldsymbol{x}))}>0
\]
because the group velocity $v_{g}(\omega)>0$.

\section{Applications}\label{section4}

\subsection{Moving source in non dispersive medias}\label{section4.1}

We suppose here that the electric and magnetic permittivity $\varepsilon$, $\mu$,
and hence the light speed $c=\frac{1}{\sqrt{\varepsilon\mu}}$ are independent
of~$\omega$. We consider a moving source of the form (\ref{1.12}) where~$A(t)$
and~$\boldsymbol{x}_0 (t)$ have form~(\ref{s1}),~(\ref{s2}). In this case
\[
S(t,\boldsymbol{x},\omega,\tau)=\frac{\omega\left\vert \boldsymbol{x}-\boldsymbol{x}_0
(\tau)\right\vert }{c}-\omega(t-\tau)-\omega_0 \tau.
\]

System (\ref{1.25}) accepts the form
\begin{gather}
\frac{\left\vert \boldsymbol{x}-\boldsymbol{x}_0 (\tau)\right\vert }{c}-(t-\tau)=0,
\qquad
-\frac{\omega}{c}v(\boldsymbol{x},\tau)+(\omega-\omega_0)=0\label{n1}
\end{gather}
and the f\/irst equation 
(\ref{n1}) independent of $\omega$. Note that under
subliminal velocity of the source
\begin{gather*}
\sup_{t}\left\vert \boldsymbol{v}(t)\right\vert <c 
\end{gather*}
f\/irst equation in (\ref{n1}) has an unique solution $\tau_{s}=\tau
_{s}(t,\boldsymbol{x})$ for every points $t$, $\boldsymbol{x}$. Second equation in
(\ref{n1}) implies that
\[
\omega_{s}=\omega_{s}(t,\boldsymbol{x})=\frac{\omega_0 }{1-\frac{v(\boldsymbol{x},
\tau_{s})}{c}}.
\]
Moreover
\begin{gather}
\det S^{\prime\prime}(t,\boldsymbol{x},\omega_{s},\tau_{s})=-\left(1-\frac{v(\boldsymbol{x},\tau_{s})}{c}\right)^{2},
\label{n3}
\end{gather}
and
\begin{gather}
\operatorname{sgn}S^{\prime\prime}(t,\boldsymbol{x},\omega_{s},\tau_{s})=0. \label{n.4}
\end{gather}
Substituting $\omega_{s}$, $\tau_{s}$, $\det S^{\prime\prime}(\omega_{s}$, $\tau
_{s})$, $\operatorname{sgn}S^{\prime\prime}(\omega_{s},\tau_{s})$ from (\ref{n3}),
(\ref{n.4}) in formulae (\ref{h2}), (\ref{h3}) we obtain the expression for
$\boldsymbol{H}(t,\boldsymbol{x})$ and $\boldsymbol{E}(t,\boldsymbol{x})$
\begin{gather*}
\boldsymbol{H}(t,\boldsymbol{x})\sim\frac{\nabla_{\boldsymbol{x}}
\times\left(\frac{e^{iS(t,\boldsymbol{x},\omega_{s},\tau_{s})}}{\left\vert
\boldsymbol{x}-\boldsymbol{x}_0 (\tau_{s})\right\vert }\boldsymbol{v}(\tau_{s})\right) a(\tau
_{s})}{4\pi\left(1-\frac{v(\boldsymbol{x},\tau_{s})}{c}\right) },
\end{gather*}
and
\begin{gather*}
\boldsymbol{E}(t,\boldsymbol{x})\sim\frac{1}{4\pi i}
\frac{a(\tau_{s})\omega_{s}\mu(\omega_{s})}{\left(1-\frac{v(\boldsymbol{x},\tau_{s})}{c}\right)}
\left(I+\frac{1}{k^{2}(\omega_{s})}\nabla_{\boldsymbol{x}}\nabla_{\boldsymbol{x}}\cdot\right)
\frac{e^{iS(t,\boldsymbol{x},\omega_{s},\tau_{s})}}{\left\vert
\boldsymbol{x}-\boldsymbol{x}_0 (\tau_{s})\right\vert }\boldsymbol{v}(\tau_{s})
\end{gather*}
for $t=\frac{T}{\lambda}$,
$\left\vert \boldsymbol{x}-\boldsymbol{x}_0 (t)\right\vert
=\frac{\left\vert \boldsymbol{X} - \boldsymbol{X}_0 (T)\right\vert }{\lambda}$,
$\lambda\rightarrow\infty$.

It should be noted that these formulae are \textit{asymptotic simplifications}
of the Li\`{e}nard--Wiechert  potentials   \cite[Chapter~14]{Jackson}.

\subsection{Modulated stationary source in dispersive media}\label{section4.2}

Let us consider the electromagnetic f\/ield generated by a \textit{modulated
stationary source }of the form
\[
\boldsymbol{j}(t,\boldsymbol{x})=A(t)e^{-i\omega_0 t}\delta(\boldsymbol{x}-\boldsymbol{x}
_0)\boldsymbol{e},\qquad \omega_0 >0, \qquad A(t)=a(t/\lambda),
\]
where $\lambda>0$ is a large parameter, $\boldsymbol{e}\in\mathbb{R}^{3}$ is a
unit vector, $\omega_0 >0$ is an eigenfrequency of the source.

Repeating the calculations carried out for obtaining formulae (\ref{1.13}),
(\ref{1.13'}) we obtain
\begin{gather*}
\boldsymbol{H}(t,\boldsymbol{x})=\frac{1}{8\pi^{2}}\int_{\mathbb{R}^{2}}
\nabla_{\boldsymbol{x}}\times\left(\frac{e^{i(k(\omega)\left\vert \boldsymbol{x}-\boldsymbol{x}
_0 \right\vert -\omega(t-\tau))}}{\left\vert \boldsymbol{x}-\boldsymbol{x}_0 \right\vert
}\boldsymbol{e}\right) A(\tau){\rm d}\omega{\rm d}\tau, \label{M1}
\\
\boldsymbol{E}(t,\boldsymbol{x})=\frac{1}{8\pi^{2}i}\int_{\mathbb{R}^{2}}
A(\tau)\omega\mu(\omega)
\left(I+k^{-2}(\omega)\nabla_{\boldsymbol{x}}\nabla_{\boldsymbol{x}}\cdot\right)                    
\left(\frac{e^{i(k(\omega)\left\vert \boldsymbol{x}-\boldsymbol{x}_0 \right\vert
-\omega(t-\tau))}}{\left\vert \boldsymbol{x}-\boldsymbol{x}_0 \right\vert }\boldsymbol{e}\right)
{\rm d}\omega{\rm d}\tau. 
\end{gather*}
The further asymptotic analyses of $\boldsymbol{H}(t,\boldsymbol{x})$, $\boldsymbol{E}
(t,\boldsymbol{x})$ is completely similar to given in Section~\ref{section4.1}. The phase~$S$
in this case is
\[
S(t,\boldsymbol{x},\omega,\tau)=k(\omega)\left\vert \boldsymbol{x}-\boldsymbol{x}_0 \right\vert
-\omega(t-\tau)-\omega_0 \tau.
\]
Hence system (\ref{1.25}) accepts the form
\begin{gather*}
\tau=t-\frac{\left\vert \boldsymbol{x}-\boldsymbol{x}_0 \right\vert }{v_{g}(\omega)},\qquad
\omega=\omega_0,
\end{gather*}
The phase $S(t,\boldsymbol{x},\omega,\tau)$ has the unique stationary
point
\[
\omega_{s}=\omega_0, \qquad \tau_{s}=t-\frac{\left\vert \boldsymbol{x}-\boldsymbol{x}_0 \right\vert
}{v_{g}(\omega_0)}
\qquad
\text{and}
\qquad
S^{\prime\prime}(t,\boldsymbol{x},\omega_0,\tau_{s})=
\begin{pmatrix}
k^{\prime\prime}(\omega_0)\left\vert \boldsymbol{x}-\boldsymbol{x}_0 \right\vert & 1\\
1 & 0
\end{pmatrix}.
\]
Hence $\det$ $S^{\prime\prime}(t,\boldsymbol{x},\omega_0,\tau_{s})=-1$,
$\operatorname{sgn}S(t,\boldsymbol{x},\omega_0,\tau_{s})=0$ and
\[
S(t,\boldsymbol{x},\omega_0,\tau_{s})=k(\omega_0)\left\vert
\boldsymbol{x}-\boldsymbol{x}_0 \right\vert -\omega_0 \tau_{s}.
\]
It implies that
\begin{gather*}
\boldsymbol{H}(t,\boldsymbol{x})\sim\frac{1}{4\pi}a\left(t-\frac{\left\vert
\boldsymbol{x}-\boldsymbol{x}_0 \right\vert }{v_{g}(\omega_0)}\right) \nabla_{\boldsymbol{x}
}\times\left(\frac{e^{i\left(k(\omega_0)\left\vert \boldsymbol{x}-\boldsymbol{x}
_0 \right\vert -\omega_0 \tau_{s}\right) }}{\left\vert \boldsymbol{x}-\boldsymbol{x}
_0 \right\vert }\boldsymbol{e}\right),
\\
\boldsymbol{E}(t,\boldsymbol{x})\sim\frac{\omega_{s}\mu(\omega_{s})a\left(t-\frac{\left\vert
\boldsymbol{x}-\boldsymbol{x}_0 \right\vert }{v_{g}(\omega_0)}\right)}{4\pi i}
\left(I+\frac{1}{k^{2}(\omega_0)}\nabla_{\boldsymbol{x}}\nabla_{\boldsymbol{x}}\cdot\right)
\frac{e^{iS(t,\boldsymbol{x},\omega_0,\tau_{s})}}{\left\vert
\boldsymbol{x}-\boldsymbol{x}_0 (\tau_{s})\right\vert }\boldsymbol{e}
\end{gather*}
for the ``large'' time and distance between the source and the receiver.
Note that the retarded time is $t-\frac{\left\vert \boldsymbol{x}-\boldsymbol{x}
_0 \right\vert }{v_{g}(\omega_0)}$.

\subsection{Propagation from a moving source in the plasma}\label{section4.3}

We consider a lossless no magnetized plasma whose the collision frequency
equals to zero (see for instance \cite{Ginz3, Stix,Swen}).
Hence the constitutive parameters in plasma are
\[
\varepsilon(\omega)=\varepsilon_0\left(1-\frac{\omega_{p}^{2}}{\omega^{2}}\right),\qquad\mu=\mu_0,
\]
$\varepsilon_0$, $\mu_0$ are the electric and magnetic permittivity of the vacuum,
\[
\omega_{p}^{2}=\frac{nq^{2}}{m\varepsilon_0 },
\]
where $\omega _{p}$ is the plasma
frequency, $n$~is the particle density, $m$, $q$ are the mass
and charge of the electron. Hence the phase velocity in the plasma is
\[
c(\omega)=\frac{c_0 }{\sqrt{1-\frac{\omega_{p}^{2}}{\omega^{2}}}},
\]
the wave-number is
\[
k(\omega)=\frac{\sqrt{\omega^{2}-\omega_{p}^{2}}}{c_0 },
\]
and the group velocity is
\[
v_{g}(\omega)=c_0 \sqrt{1-\frac{\omega_{p}^{2}}{\omega^{2}}},
\]
where $c_0 $ is the light speed in the vacuum.

We consider the electromagnetic f\/ield in the plasma generated by a moving
source of the form~(\ref{1.12}) under conditions (\ref{s1}), (\ref{s1'}),
(\ref{s2}). The phase $S$ is
\begin{gather*}
S(t,\boldsymbol{x},\omega,\tau)=\frac{\sqrt{\omega^{2}-\omega_{p}^{2}}}{c_0
}\left\vert \boldsymbol{x}-\boldsymbol{x}_0 (\tau)\right\vert -\omega(t-\tau)-\omega_0 \tau.
\end{gather*}
We suppose that $\omega_0 >\omega_{p}$. System (\ref{1.25}) accepts the form
\begin{gather}
\frac{\left\vert \boldsymbol{x}-\boldsymbol{x}_0 (\tau)\right\vert }{c_0 \sqrt{1-\frac
{\omega_{p}^{2}}{\omega^{2}}}}-(t-\tau)=0,
\qquad
-\frac{\sqrt{\omega^{2}-\omega_{p}^{2}}}{c_0}v(\boldsymbol{x},\tau)+(\omega-\omega_0)=0,\label{1.40}
\end{gather}
and under the condition
\begin{gather*}
\sup_{t,\omega\geq\omega_{p}}\frac{\left\vert \boldsymbol{v}(t)
\right\vert }{v_{g}(\omega)}<1 
\end{gather*}
system (\ref{1.40}) has an unique solution $(\omega_{s},\tau_{s})$ such that
$\omega_{s}>\omega_{p}$.

The substitution $\left(\omega_{s},\tau_{s}\right) $ in formulae
(\ref{h2}), (\ref{h3}) gives the expression for the electromagnetic f\/ield
generated by the moving source.

\begin{example}
Let $\boldsymbol{v}$ be a constant vector and let $v(\boldsymbol{x},t)=\pm\left\vert
\boldsymbol{v}\right\vert $. In this case equations (\ref{1.40}) accept the form
\begin{gather}
\tau=t-\frac{\left\vert \boldsymbol{x}-\boldsymbol{x}_0 (\tau)\right\vert }{c_0
\sqrt{1-\frac{\omega_{p}^{2}}{\omega^{2}}}},
\qquad
\omega=\omega_0 \pm\frac{\sqrt{\omega^{2}-\omega_{p}^{2}}}{c_0
}\left\vert \boldsymbol{v}\right\vert,\label{1.41}
\end{gather}
where the sign $+$ is taken if the source moving to the receiver and the sign
$-$ if the source moving from the receiver. We obtain from second equation in
(\ref{1.41}) that
\[
\omega_{s}^{\pm}=\frac{1}{1-M^{2}}\left(\omega_0 \pm M\sqrt{\omega_0
^{2}-(1-M^{2})\omega_{p}^{2}}\right),
\]
where $M=\frac{\left\vert \boldsymbol{v}\right\vert }{c_0 }$ $(<1)$ is the Mach
number. For $\omega=\omega_{s}^{\pm}$ f\/irst equation in (\ref{1.41}) has the
unique solution $\tau_{s}^{\pm}$. It is easy to see that
\begin{gather}
\det S^{\prime\prime}(t,\boldsymbol{x},\omega_{s}^{\pm},\tau_{s}^{\pm})=-\left(1\pm\frac{\left\vert
\boldsymbol{v}\right\vert }{c_0 \sqrt{1-\frac{\omega_{p}
^{2}}{(\omega_{s}^{\pm})^{2}}}}\right)^{2}, \label{1.42}
\end{gather}
and
\begin{gather}
\operatorname{sgn}S^{\prime\prime}(t,\boldsymbol{x},\omega_{s}^{\pm},\tau_{s}^{\pm})=0.
\label{1.43}
\end{gather}
The substitution $\left(\omega_{s},\tau_{s}\right) $ and (\ref{1.42}),
(\ref{1.43}) in formulae (\ref{h2}), (\ref{h3}) gives the expressions for the
electromagnetic f\/ield
\begin{gather*}
\boldsymbol{E}_{\pm}(t,\boldsymbol{x})\sim\frac{1}{4\pi i}
\frac{a(\tau_{s}^{\pm})\omega_{s}^{\pm}\mu}
{\left(1\pm\frac{\left\vert \boldsymbol{v}\right\vert}
{c_0 \sqrt{1-\frac{\omega_{p}^{2}}{(\omega_{s}^{\pm})^{2}}}}\right)}
\left(I+\frac{c_0 ^{2}}{\left(\omega_{s}^{\pm}\right)^{2}
-\omega_{p}^{2}}\nabla_{\boldsymbol{x}}\nabla_{\boldsymbol{x}}\cdot\right)
\frac{e^{iS(t,\boldsymbol{x},\omega_{s}^{\pm},\tau_{s}^{\pm})}}{\left\vert
\boldsymbol{x}-\boldsymbol{x}_0 (\tau_{s})\right\vert }\boldsymbol{v} ,
\\
\boldsymbol{H}_{\pm}(t,\boldsymbol{x}\sim\frac{1}{4\pi}\nabla_{\boldsymbol{x}}
\times\left(\frac{e^{iS(t,\boldsymbol{x},\omega_{s}^{\pm},\tau_{j})}}{\left\vert
\boldsymbol{x}-\boldsymbol{x}_0 (\tau_{j})\right\vert }\boldsymbol{v}\right) \frac{a(\tau
_{s}^{\pm})}{\left(1\pm\frac{\left\vert \boldsymbol{v}\right\vert }{c_0
\sqrt{1-\frac{\omega_{p}^{2}}{(\omega_{s}^{\pm})^{2}}}}\right) },
\end{gather*}
for $t=\frac{T}{\lambda}$, $\left\vert \boldsymbol{x}-\boldsymbol{x}_0 (t)\right\vert
=\frac{\left\vert \boldsymbol{X} - \boldsymbol{X}_0 (T)\right\vert}{\lambda}$, $\lambda\rightarrow\infty$.
\end{example}

\subsection{Cherenkov radiation}\label{section4.4}

Now we apply the above developed approach to consider a f\/ield radiation from a
moving with a~constant velocity $\boldsymbol{v}$ charged particle. Here $e$ is a
particle charge, while $c(\omega)>0$ is a f\/ield phase velocity in the
isotropic dispersive medium. Following \cite[Chapter~XIV]{LL8}
we use
\begin{gather*}
\rho(t,\boldsymbol{x})=e\delta(\boldsymbol{x}-\boldsymbol{v}t),
\qquad
\boldsymbol{j}(t,\boldsymbol{x})=\boldsymbol{v}e\delta(\boldsymbol{x}-\boldsymbol{v}t).
\end{gather*}
We suppose that $\boldsymbol{v}=(0,0,v)$ and $\boldsymbol{x}=(x_{1},x_{2},x_{3})$. For
the particle eigenfrequency case $\omega_0 =0$ the phase $S$ reads
\[
S(t,\boldsymbol{x},\omega,\tau)=k(\omega)\left\vert \boldsymbol{x}-\boldsymbol{v}
t\right\vert -\omega(t-\tau).
\]
Hence system (\ref{1.25}) accepts the form
\begin{gather}
\frac{\partial S(t,\boldsymbol{x},\omega,\tau)}{\partial\tau}   =-\frac{\omega
}{c(\omega)}v(\boldsymbol{x},\tau)+\omega=0,\label{ch1}\\
\frac{\partial S(t,\boldsymbol{x},\omega,\tau)}{\partial\omega}   =\frac
{\left\vert \boldsymbol{x}-\boldsymbol{v}\tau\right\vert }{v_{g}(\omega)}- (t-\tau)=0.
\label{ch1'}
\end{gather}
We suppose that the system (\ref{ch1}), (\ref{ch1'}) has a solution
$(\omega_{s},\tau_{s})$ with a non-trivial frequency $\omega_{s}>0$. So, is
there exists a pair $(\omega_{s},\tau_{s})$ such that
\begin{gather}
v(\boldsymbol{x},\tau_{s})=\left\vert \boldsymbol{v}\right\vert \cos\varphi
(\boldsymbol{x},\tau_{s})=c(\omega_{s})>0, \label{ch2'}
\end{gather}
where $\varphi(\boldsymbol{x},\tau)$ is the angle between $\boldsymbol{v}$ and
$\boldsymbol{x}-\boldsymbol{v}\tau$. The equation~(\ref{ch2'}) can be satisf\/ied if the value of the
projection of the velocity $\boldsymbol{v}$ on the vector $\boldsymbol{x}-\boldsymbol{v}t$ is positive.

Let now the pair $(\omega_{s},\tau_{s})$
be an isolated non-degenerated
stationary point of the phase $S$,
\begin{gather}
v(\boldsymbol{x},\tau_{s})=c(\omega_{s}),\label{ch2}\\
\tau_{s}=t-\frac{\left\vert \boldsymbol{x}-\boldsymbol{v}\tau_{s}\right\vert}{v_{g}(\omega_{s})}, \label{ch3}
\end{gather}
and
$
\det S^{\prime\prime}(t,\boldsymbol{x},\omega_{s},\tau_{s})\neq0
$.

Accordingly to the causality principle the root $\tau_{s}$ of the equation
(\ref{ch3}) has to be positive. It implies the Cherenkov cone condition
\begin{gather}
vt-x_{3}-\left\vert \boldsymbol{x}^{\prime}\right\vert \sqrt{\left\vert \beta
^{2}(\omega_{s})-1\right\vert }>0, \label{ch4}
\end{gather}
where $\boldsymbol{x}^{\prime}=(x_{1},x_{2})$ and
\begin{gather*}
\beta(\omega_{s})=\frac{v}{v_{g}(\omega_{s})} 
\end{gather*}
(see, e.g., \cite{Afanasiev:2004a} and references therein). Condition
(\ref{ch4}) connects the time $t$ and position $\boldsymbol{x}=(x_{1},x_{2}
,x_{3})$ of the point where the Cherenkov radiation exists.

The substitution of $(\omega_{s},\tau_{s})$ to equations~(\ref{h2}),~(\ref{h3})
leads to the following expressions for the electromagnetic f\/ields
$\boldsymbol{E}_{s}(t,\boldsymbol{x})$, $\boldsymbol{H}_{s} (t,\boldsymbol{x})$ (at the point
$(t,\boldsymbol{x})$) emitted by the moving charge with the instantaneous
frequency $\omega_{s}=\omega_{s}(t,\boldsymbol{x})>0$ (the Cherenkov
radiation)
\begin{gather*}
\boldsymbol{H}_{s}(t,\boldsymbol{x})
\sim\frac{1}{4\pi}\nabla_{\boldsymbol{x}}
\times\left(\frac{e^{iS(t,\boldsymbol{x},\omega_{s},\tau_{s})}
}{\left\vert \boldsymbol{x}-\boldsymbol{v}\tau_{s})\right\vert }\boldsymbol{v}\right)
\frac{e^{\frac{i\pi}{4}\operatorname{sgn}S^{\prime\prime}(t,\boldsymbol{x},\omega_{s}
,\tau_{s})}}{\left\vert \det S^{\prime\prime}(t,\boldsymbol{x},\omega
_{s},\tau_{s})\right\vert^{1/2}}\nonumber\\
\phantom{\boldsymbol{H}_{s}(t,\boldsymbol{x})\sim }
\times\Theta\Big(vt-x_{3}- \vert \boldsymbol{x}^{\prime} \vert
\sqrt{\left\vert \beta^{2}(\omega_{s})-1\right\vert}\Big), 
\\
\boldsymbol{E}_{s}(t,\boldsymbol{x})
\sim\frac{1}{4\pi i}\omega_{s}
\mu(\omega_{s})(I+\frac{1}{k^{2}(\omega_{s})}\nabla_{\boldsymbol{x}}
\nabla_{\boldsymbol{x}}\cdot)\frac{e^{iS(t,\boldsymbol{x},\omega_{s},\tau
_{s})}}{\left\vert \boldsymbol{x}-\boldsymbol{v}\tau_{s}\right\vert }\boldsymbol{v}\nonumber\\
\phantom{\boldsymbol{E}_{s}(t,\boldsymbol{x})\sim}
\times\frac{e^{\frac{i\pi}{4}\operatorname{sgn}S^{\prime\prime}(t,\boldsymbol{x},\omega
_{s},\tau_{s})}}{\left\vert \det S^{\prime\prime}(t,\boldsymbol{x},
\omega_{s},\tau_{s})\right\vert^{1/2}}\Theta\Big(vt-x_{3}- \vert
\boldsymbol{x}^{\prime} \vert \sqrt{\left\vert \beta^{2}(\omega
_{s})-1\right\vert }\Big),
\end{gather*}
where $\Theta(r)$ is the Heaviside function that is
\[
\Theta(r)=
\begin{cases}
1, & r>0,\\
0, & r\leq0.
\end{cases}
\]

\section{Doppler ef\/fect in metamaterials. Numerical example}\label{section5}

Further investigations of the f\/ield properties in a dispersive medium already
requires the know\-ledge of the spectral properties of the material refraction
index. So, for the further we have to choose the type of material dispersion,
in recent literature normally the Lorenz or Drude models, see
\cite{Grzegorczyk:2003a,Duan:2009a} and references therein. Never the
less, it is still dif\/f\/icult to investigate the properties of the
electromagnetic waves in such dispersive model analytically. (We note that
recently some semi-analytic methods were developed~\cite{Vorobev:2012a}).
As the example of the developed approach, below we apply the numerics to study
the spectral properties of the Doppler ef\/fect in a dispersive medium. Here we
concentrate on the dispersive metamaterials case where the refraction index
$n$ can be negative $\operatorname{Re}(n)<0$ (NIM). (In literature also refer to such a
material as being left-handed (LH) material). We consider such a medium
characterized by a (relative) permittivity $\varepsilon(\omega)$ and a
(relative) permeability $\mu(\omega)$, both of which are complex functions of
frequency $\omega$ and the refraction index $n(\omega)$ satisfying the
relations \cite{TrungDung:2003b,Ziolkowski01}
\begin{gather}
n(\omega)=\sqrt{|\varepsilon(\omega)\mu(\omega)|} e^{i[\phi_{\varepsilon
}(\omega)+\phi_{\mu}(\omega)]/2}. \label{indRefrNIM}
\end{gather}
In order to allow a frequency dependence of the refractive index $n$, let us
restrict our attention to a single-resonance permittivity
\begin{gather}
\varepsilon(\omega)=1+\frac{\omega_{{\rm P}e}^{2}}{\omega_{{\rm T}e}
^{2}-\omega^{2}-i\omega\gamma_{e}} \label{eps}
\end{gather}
and a single-resonance permeability
\begin{gather}
\mu(\omega)=1+\frac{\omega_{{\rm P}m}^{2}}{\omega_{{\rm T}m}^{2}
-\omega^{2}-i\omega\gamma_{m}} , \label{mu}
\end{gather}
where $\omega_{{\rm P}e}$, $\omega_{{\rm P}m}$ are the coupling
strengths, $\omega_{{\rm T}e}$, $\omega_{{\rm T}m}$ are the transverse
resonance frequencies, and $\gamma_{e}$, $\gamma_{m}$ are the absorption
parameters. Both the permittivity and the permeability satisfy the
Kramers--Kronig  relations \cite{TrungDung:2003b, Ziolkowski01}. The
following typical parameters of metamaterial were used for our numerics:
$f_{{\rm P}e}=298.42$~THz, $\gamma_{e}=0.04$~THz , $f_{{\rm T}e}=409.82$~THz, $f_{{\rm P}m}
=171.09$~THz, $\gamma_{m}=0.04$~THz, $f_{{\rm T}m}=397.89$~THz.

\begin{figure}[t]
\centering
\includegraphics[scale=0.60]{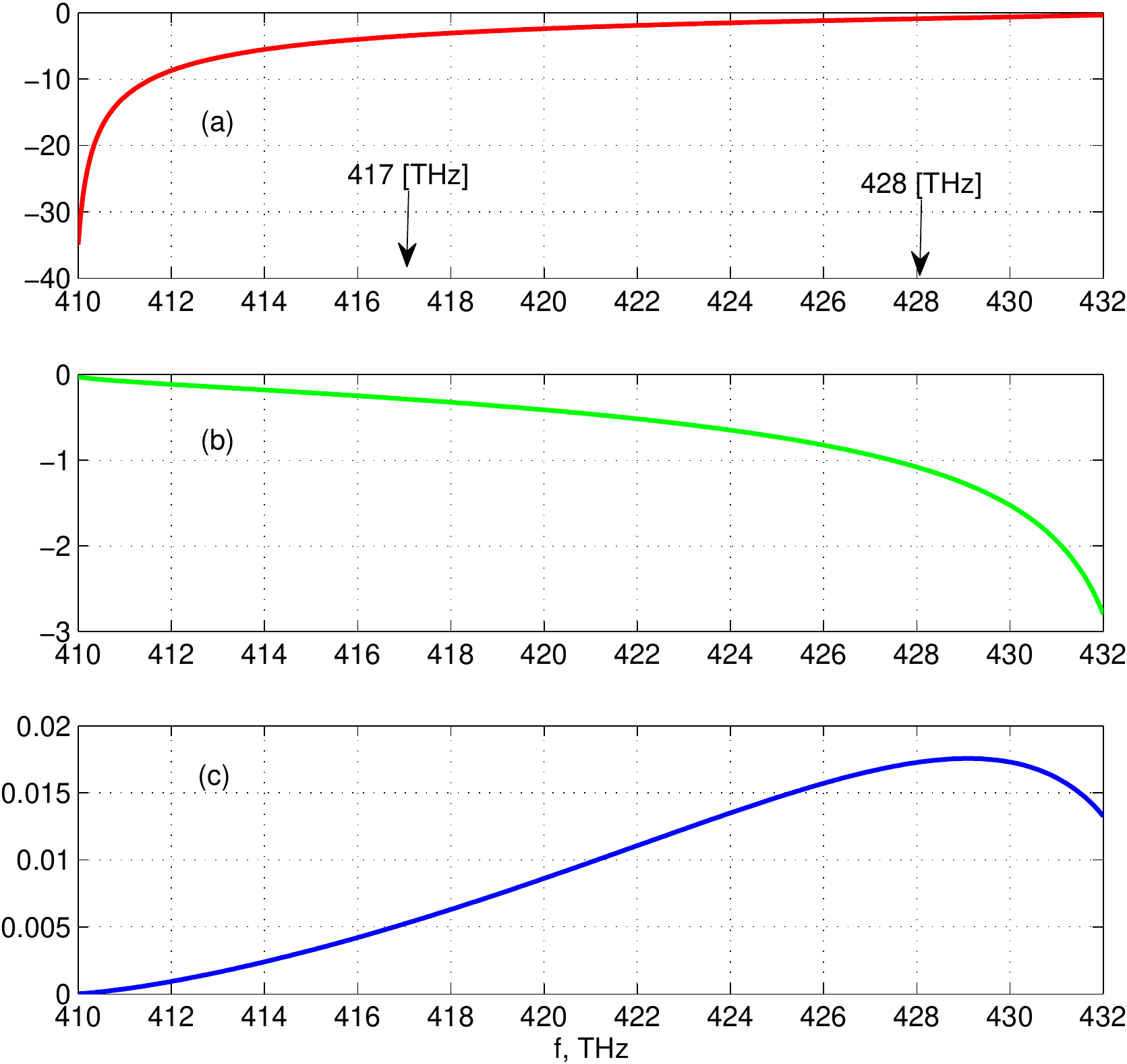}
\caption{(Color online.) Frequency dependence (a) metaterial refractive index
$n(\omega)$ in the frequency interval with $\operatorname{Re}n(\omega)<0$; (b)
phase velocity $v_{p}(\omega)$; (c) group velocity $v_{g}(\omega)$, where
$\omega=2\pi f$.}
\label{Pic_fig1}
\end{figure}

Fig.~\ref{Pic_fig1}(a) shows the frequency dependence of NIM refractive index
$n(\omega)$ ($\omega=2\pi f$). Here the permittivity $\varepsilon(\omega)$ and
the permeability $\mu(\omega)$ being respectively given by equations~(\ref{eps})
and~(\ref{mu}) in the frequency interval from $410$~THz to $432$~THz where
$\operatorname{Re}n(\omega)<0$. It is worth noting that the negative real part
of the refractive index is typically observed together with strong dispersion,
so that absorption cannot be disregarded in general. However, in a recent
experiment~\cite{Xiao:2010a}, it was demonstrated that the incorporation of
gain material in a metamaterial makes it possible to fabricate an extremely
low-loss and active optical devices. Thus, the original loss-limited negative
refractive index can be drastically improved with loss compensation in the
visible wavelength range. Following this result we will neglect the imaginary
part $\operatorname{Im}(n)$. Fig.~\ref{Pic_fig1}(b) shows corresponding
frequency dependence of the phase velocity $v_{p}(\omega)$ that is negative in
considered frequency range, and Fig.~\ref{Pic_fig1}(c) exhibits the frequency
dependence of the group velocity $v_{g}(\omega)$. Further for numerics we
renormalized the variables for $\tau$ and $x$ with the scales $l_0 /c$ and
$l_0 $ respectively, velocities are normalized with the vacuum light velocity~$c$, and $l_0 =75$~nm is the typical spatial scale used at the metamaterial
experiments~\cite{Xiao:2010a}. To seek for simplicity, further we consider with
details a case when the position of observer $\boldsymbol{x}$ and the trajectory
of a~source $\boldsymbol{v}$ are in the same plane ($x_{3}=0$ and $H=0$). In this
case the geometry becomes 2D one and the solution to equations~(\ref{1.25}) for
$\omega$ reads
\begin{gather}
\omega=\left(c{r}^{2}\pm\sqrt{{v}^{2}{r}^{2}\left(-x_{2}+v\tau\right)
^{2}n(\omega)^{2}}\right) \frac{c\omega_0 }{{c}^{2}{r}^{2}-p_{2}},
\label{w 2D}
\end{gather}
where $p_{2}=n(\omega)^{2}{v}^{2}\left(x_{2}-v\tau\right)^{2}$, and
$r=\sqrt{x_{1}^{2}+\left(x_{2}-v\tau\right)^{2}}$. The solution of
equations~(\ref{1.25}) for~$\tau$ can be written as
\begin{gather}
\tau= \frac{x_{2} v-v_{g}^{2}t-\sqrt{q_{1}+q_{2}}}{{v}^{2}-v_{g}^{2}},
  \label{tau 2D}
\end{gather}
where $q_{1}=\left(v^{2}-v_{g}^{2}\right) x_{1}^{2}$, and $q_{2}=v_{g}
^{2}\left(x_{2}-vt\right)^{2}$. In simplest 1D situation with $x_{1}=0$
from~(\ref{w 2D}),~(\ref{tau 2D}) we have two equations
\begin{gather}
\omega={\frac{\omega_0 }{1\pm n(\omega)v}},\qquad \tau={\frac{x_{2}-v_{g}
t}{v-v_{g}}}. \label{w 1D}
\end{gather}
First formula in equation~(\ref{w 1D}) describes the well-known Doppler ef\/fect, when
the shift of $\omega$ depends on the~$nv$ signum. For conventional materials
with $n>0$ (in this case we have to choose upper $+$ sign) the received
frequency $\omega$ is lower (compared to the emitted frequency~$\omega_0 $)
during the source approach $v>0$, and $\omega$ is higher during the source
recession $v<0$. However the situation becomes more complicated in dispersive
metamaterials with negative refraction index (NIM) where $\operatorname{Re}
(n(\omega))<0$ with the refraction index depending on the frequency
$n=n(\omega)$ were both phase and group velocities are the frequency
functions. In dispersive medium the f\/irst relation in~(\ref{w 2D}) and~(\ref{tau 2D}) become coupled equations. Even in a simple 1D geometry the
f\/irst equation for frequency~$\omega$ in~(\ref{w 1D}) already requires the
spectral details of the medium refraction index~$n(\omega)$. Moreover, in the
second equation in~(\ref{w 1D}) the group velocity~$v_{g}$ becomes the
frequency function, thus, the retardation time $\tau$ will be dif\/ferent for
dif\/ferent source eigenfrequency~$\omega_0 $.

\begin{figure}[t]
\centering
\includegraphics[scale=0.60]{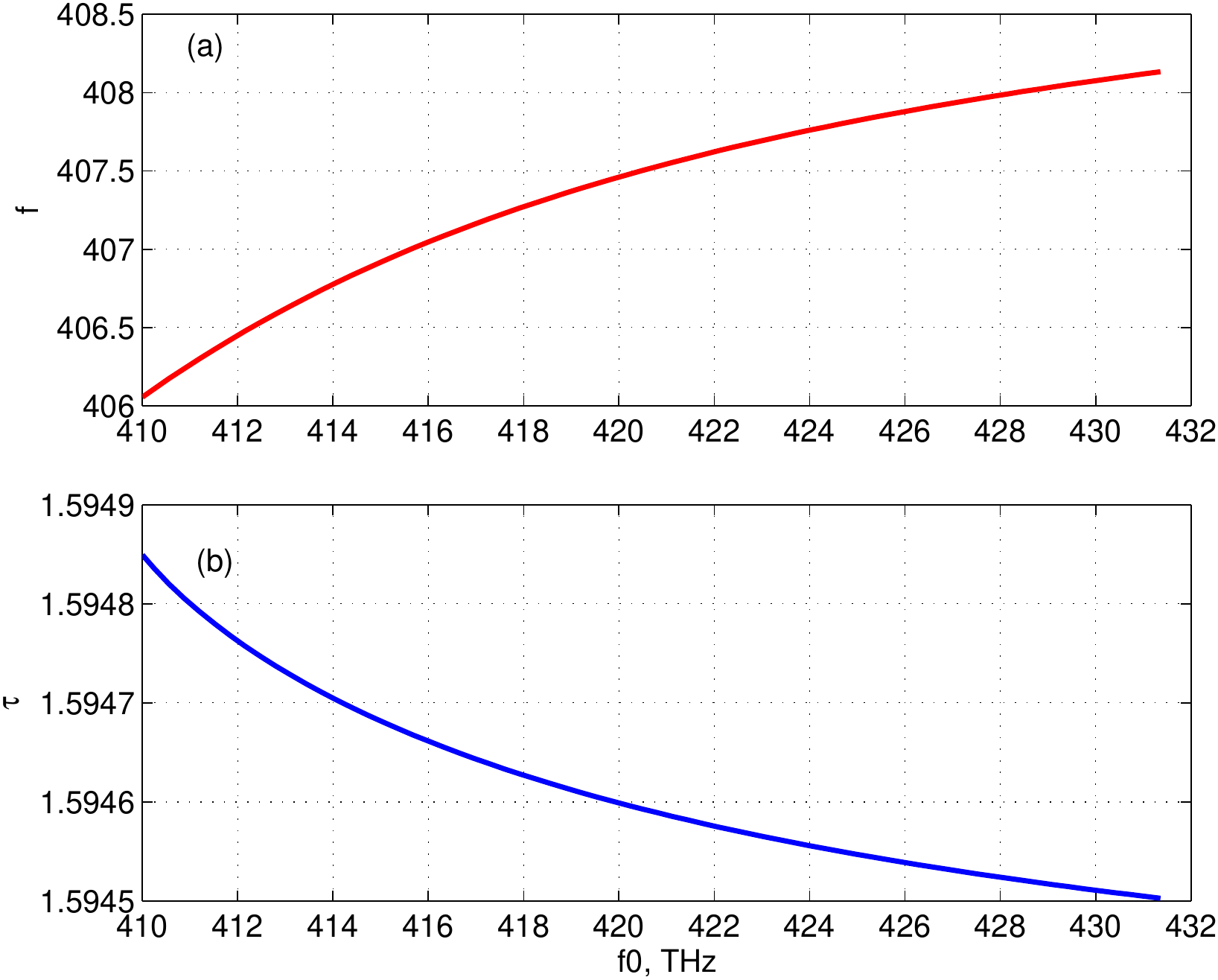}
\caption{(Color online.) Dependence of the Doppler shifted frequency $\omega$
($\omega=2\pi f$)~(a), and the time~$\tau$~(b) on the source frequency
$\omega_0 $ for area where~$n(\omega)<0$. See details in text.}
\label{Pic_fig2}
\end{figure}

In our numerical simulations we search for the solution of equations
(\ref{indRefrNIM})--(\ref{w 1D}) for $\omega$  and $\tau$  at the f\/ixed
position $x_{1}$, $x_{2}$ and the time $t$ of an observer. First we studied
more simple 1D dispersive NIM, equation~(\ref{w 1D}). This equation is solved by
the standard numerical methods\cite{Press,AllenTaflove:2005a}, and resulting dependencies ares
shown in Fig.~\ref{Pic_fig2}. Fig.~\ref{Pic_fig2}(a) shows the shifted
frequency $\omega$ ($\omega=2\pi f$) vs the source frequency $\omega_0 $ for
area where $n(\omega)<0$. Further such a calculated dependence $\omega
=\omega(\omega_0)$ allows us to evaluate the group velocity $v_{g}(\omega)$
and, f\/inely the retardation time $\tau=\tau(\omega_0)$ in Fig.~\ref{Pic_fig2}(b). We observe from Fig.~\ref{Pic_fig2}(a),~(b) that both dependencies
$\omega=\omega(\omega_0)$ and $\tau=\tau(\omega_0)$ have pronounced
nonlinear shape that certainly is determined by a dispersive spectrum of used
metamaterial. Let us remind that in dispersiveless case both Dopple's
frequency shift and the retardation time depend on the particle velocity $v$ only.

The 2D geometry is more complicated. In this situation we have to solve
numerically the equations (\ref{w 2D}), (\ref{tau 2D}) (with added material
relations (\ref{indRefrNIM})--(\ref{mu})) that  becomes a strong nonlinear
system. It is worth to note that in this case the Newton's method for solving
nonlinear equations has an unfortunate tendency to wander of\/f~\cite{Press,AllenTaflove:2005a} if
the initial guess is not suf\/f\/iciently close to the root. In order to evaluate
the solution of such a system the globally convergent multi-dimensional
Newton's method was applied~\cite{Press,AllenTaflove:2005a}. The radiated source has $v=0.5$, and
$f_0 =420$~THz. The observer point is at $x_{1}=0.01$, $x_{2}=1.595$, and the
time is $t=2$. The result of calculations is $f=417.82$~THz, and the
$\tau=3.1901$. This point is indicated in Fig.~\ref{Pic_fig1}(a) with
$f=417.82$~THz, and corresponds to $v_{p}=-0.31673$, $v_{g}=0.0084029$, and
$x_{2}<v\tau$. Other solution was obtained for parameters $x_{1}=0$,
$x_{2}=1.595$ that results $f=428.9$~THz and $\tau=3.1694$ (see arrows in
Fig.~\ref{Pic_fig1}(a)).

\section{Conclusion}\label{section6}

In this paper the time-frequency integrals and the two-dimensional stationary
phase method are applied to study the electromagnetic waves radiated by moving
modulated sources in dispersive media. We show that such unif\/ied approach
leads to explicit expressions for the f\/ield amplitudes and simple relations
for the f\/ield eigenfrequencies and the retardation time that become the
coupled variables. The main features of the technique are illustrated by
examples of the moving source f\/ields in the plasma and the Cherenkov
radiation. In the paper it is emphasized that deeper comprehension and insight
the wave ef\/fects in dispersive case already requires the explicit formulation
of the dispersive material model. As the advanced application we have
considered the Doppler frequency shift in a complex single-resonant dispersive
metamaterial (Lorenz) model where in some frequency ranges the negativity of
the real part of the refraction index can be reached. We have demonstrated
that in a dispersive case the Doppler frequency shift acquires a nonlinear
dependence on the modulating frequency of the radiated particle.
The detailed frequency dependence of such a shift and spectral behavior of phase and group velocities
(that have the opposite directions) are studied numerically. Such dependence in principle can be used also
for reconstruction of unknown parameters of the dispersive medium. Other challenge is to enforce the
developed approach to the Cerenkov radiation and its peculiar features in the
dispersive and left-handed media. Such a problem will be considered in future study.

\subsection*{Acknowledgements}

The work of authors is partially supported by PROMEP, grant Redes CA
2011--2012. The work of G.B.\ is partially supported by CONACyT grant 169496. The
work of V.R.\ was partially supported by CONACyT grant 179872.

\pdfbookmark[1]{References}{ref}
\LastPageEnding

\end{document}